\documentclass[aps,a4paper,superscriptaddress,nofootinbib,twocolumn]{revtex4-1}
\usepackage{amsmath,amssymb,graphicx,multirow,dcolumn,bm,latexsym,soul,nicefrac}
\usepackage[colorlinks,linkcolor=blue,citecolor=blue,urlcolor=blue ]{hyperref}
\usepackage{acronym}
\usepackage{subfigure}
\usepackage{appendix}
\usepackage{longtable}
\usepackage{footnote}
\newcommand{\be}{\begin{equation}}
\newcommand{\ee}{\end{equation}}
\newcommand{\bea}{\begin{eqnarray}}
\newcommand{\eea}{\end{eqnarray}}
\newcommand{\nn}{\nonumber}

\newcommand{\td}{\tilde}

\newcommand{\vb}{{verification binaries}}
\newcommand{\TSS}{TianQin Research Center for Gravitational Physics and School of Physics and Astronomy, Sun Yat-sen University (Zhuhai Campus), Zhuhai 519082, People's Republic of China}
\newcommand{\HUST}{MOE Key Laboratory of Fundamental Physical Quantities Measurements,
Hubei Key Laboratory of Gravitation and Quantum Physics, PGMF and School of Physics,
Huazhong University of Science and Technology, Wuhan 430074, People's Republic of China}
\def\aj{Astronomical Journal}    
\def\apj{Astrophysical Journal}  
\def\apjl{Astrophysical Journal} 

\def\mnras{MNRAS}                
\def\prd{Phys.~Rev.~D}           
\def\prl{Phys.~Rev.~Lett}        
\def\araa{ARA\&A}                
\def\nat{Nature}                 
\def\aap{A\&A}                   

\allowdisplaybreaks
\begin{document}
\title{Science with the TianQin Observatory: Preliminary results on Galactic double white dwarf binaries}

\author{Shun-Jia Huang}
\affiliation{\TSS}

\author{Yi-Ming Hu}
\email{huyiming@mail.sysu.edu.cn}
\affiliation{\TSS}

\author{Valeriya Korol}
\affiliation{School of Physics and Astronomy, University of Birmingham, Birmingham B15 2TT, United Kingdom}

\author{Peng-Cheng Li}
\affiliation{Center for High Energy Physics, Peking University, No. 5 Yiheyuan Road, Beijing 100871, People's Republic of China}
\affiliation{Department of Physics and State Key Laboratory of Nuclear Physics and Technology, Peking University, No. 5 Yiheyuan Road, Beijing 100871, People's Republic of China}
\affiliation{\TSS}

\author{Zheng-Cheng Liang}
\affiliation{\HUST}
\affiliation{\TSS}

\author{Yang Lu}
\affiliation{\TSS}

\author{Hai-Tian Wang}
\affiliation{Purple Mountain Observatory, Chinese Academy of Sciences, Nanjing 210023, People's Republic of China}
\affiliation{School of Astronomy and Space Science, University of Science and Technology of China, Hefei, Anhui 230026, People's Republic of China}
\affiliation{\TSS}

\author{Shenghua Yu}
\affiliation{National Astronomical Observatories, Chinese Academy of Sciences, Beijing 100012, People's Republic of China}
\affiliation{The Key Laboratory of Radio Astronomy, Chinese Academy of Sciences, Beijing 100012, People's Republic of China}

\author{Jianwei Mei}
\email{meijw@sysu.edu.cn}
\affiliation{\TSS}


\date{\today}

\begin{abstract}
We explore the prospects of detecting Galactic double white dwarf (DWD) binaries with the space-based gravitational wave (GW) observatory TianQin.
In this work, we analyze both a sample of currently known DWDs and a realistic synthetic population of DWDs to assess the number of guaranteed detections and the full capacity of the mission. 
We find that TianQin can detect {\bf 12} out of $\sim100$ known DWDs; GW signals of these binaries can be modeled in detail ahead of the mission launch, and therefore they can be used as verification sources.  
Besides, we estimate that TianQin has a potential to detect as many as $10^4$ DWDs in the Milky Way. 
TianQin is expected to measure their orbital periods and amplitudes with accuracies of $\sim10^{-7}$ and $\sim0.2$, respectively, and to localize on the sky a large fraction (39\%) of the detected population to better than 1 deg$^2$. 
We conclude that TianQin has the potential to significantly advance our knowledge on Galactic DWDs by increasing the sample up to 2 orders of magnitude, and will allow their multi-messenger studies in combination with electromagnetic telescopes. 
We also test the possibilities of different configurations of TianQin: (1) the same mission with a different orientation,  (2) two perpendicular constellations combined into a network, and (3) the combination of the network with the ESA-led Laser Interferometer Space Antenna.
We find that the network of detectors boosts the accuracy on the measurement of source parameters by 1-2 orders of magnitude, with the improvement on sky localization being the most significant.
	
\end{abstract}

\maketitle
\acrodef{GW}{gravitational wave}
\acrodef{EM}{electromagnetic}
\acrodef{DWD}{double white dwarf binarie}
\acrodef{WD}{white dwarf}
\acrodef{SNR}{signal-to-noise ratio}
\acrodef{FIM}{Fisher information matrix}
\acrodef{BBH}{binary black hole}
\acrodef{BNS}{binary neutron star}
\acrodef{AGN}{active galactic nuclei}
\acrodef{EMRI}{extreme mass ratio inspiral}
\acrodef{SNe Ia}{type-Ia supernovae}
\acrodef{PSD}{power spectral density}
\acrodef{CVBs}{candidate verification binaries}
\acrodef{GR}{general relativity}
\section{Introduction}\label{sec:1}

The first direct detection of \acp{GW} generated from a binary black hole merger (GW150914) was made by the LIGO and Virgo Collaborations in 2015 \cite{GW150914:2016}, one hundred years after they were predicted by Albert Einstein \cite{1916SPAW.......688E}.
This detection, together with several subsequent ones including a binary neutron star merger (GW170817), started new fields of \ac{GW} and multi-messenger astronomy \cite{PhysRevX.6.041015,PhysRevX.9.031040,2020arXiv200408342T,2020ApJ...892L...3A,2020ApJ...896L..44A}. 

The sensitivity band of the currently operational ground-based detectors LIGO and Virgo is limited between 10\,Hz and kilohertz frequencies \cite{McWilliams:2019}.
However, \ac{GW} sources span many orders of magnitude in frequency down to femtohertz.
Several experiments aim to cover such a large spectrum: the cosmic microwave background polarization experiments \citep{Kamionkowski:2016}, the pulsar timing array \citep{Arzoumanian:2018,Shannon:2015}, and the space-based laser interferometers, sensitive to femtohertz, nanohertz and millihertz frequencies, respectively \citep{LISA:2017,Luo:2015}. 

The millihertz frequency band is populated by a large variety of \ac{GW} sources:
massive black hole binaries ($10^3$ - $10^{7} \mathrm{M}_\odot$) formed via galaxy mergers \cite{Klein:2016,Wang:2019,Magorrian:1998,Lynden:1969,2019PhRvD..99l3002F}; compact stellar objects orbiting massive black holes, called extreme mass ratio inspirals (EMRIs) \cite{Babak:2017,Fan:2019};
ultra-compact stellar mass binaries (and multiples) composed of white dwarfs, neutron stars and stellar-mass black holes in the neighborhood of the Milky Way \cite{Lamberts:2018,Lau:2020,Korol:2017,Robson:2018}.
Besides individually resolved binaries, stochastic backgrounds of astrophysical and cosmological origin can be detected at millihertz frequencies \citep[e.g.][]{Romano:2017,Liang:2019}. 
Therefore, this band is expected to provide rich and diverse science, ranging from Galactic astronomy to high-redshift cosmology and to fundamental physics \cite{Korol:2018b,Berti:2005,Shi:2019,2019PhRvD.100h4024B,Tamanini:2018}.

Among all kinds of ultra-compact stellar mass binaries, those composed of two white dwarf stars  [\acp{DWD}] comprise the absolute majority (up to $10^8$) in the Milky Way. Being abundant and nearby, \acp{DWD} are expected to be the most numerous \ac{GW} sources for space-based detectors \cite{Nelemans:2001b,Yu:2010,Breivik:2019,Lamberts:2018}.

Individual \ac{GW} detections of \acp{DWD} will significantly advance our knowledge on binary formation and white dwarf stars themselves in a number of ways.
First, \acp{DWD} represent the end products of the low-mass binary evolution, and as such they encode information on physical processes such as the highly uncertain mass transfer and common envelope phases \cite{Postnov:2014,Belczynski:2002}.
Second, \acp{DWD} are progenitors to AM canum venaticorum (AM CVn) systems, short-period ($\lesssim$1 hour) mass-transferring \acp{DWD}, ideal for studying the stability of the mass transfer \cite{Nelemans:2001a,Marsh:2004,Solheim:2010,2018PhRvL.121m1105T}.
Third, \ac{DWD} mergers are thought to originate a broad range of interesting transient events including \ac{SNe Ia} \cite{Bildsten:2007,Webbink:1984,Iben:1984,Webbink:1984}. 
In addition, detached \acp{DWD} are particularly suitable for studying the physics of tides. \acp{DWD} affected by tides will yield information on the nature and origin of white dwarf viscosity, which is still a missing piece in our understanding of white dwarfs' interior matter \cite{Piro:2011,Fuller:2012,Dall'Osso:2014,mckernan:2016}.
Finally, by analyzing their \ac{GW} signals one could set constraints on  deviations from general relativity \cite{Littenberg:2018,Cooray:2004}.

The overall \ac{GW} signal from \acp{DWD} imprints the information on the Galactic stellar population as a whole, and it can constrain the structural properties of the Milky Way \cite{Benacquista:2006,Adams:2012,Korol:2018,Breivik:2019,Wilhelm:2020}. 
A significant fraction of the population may present a stellar or sub-stellar tertiary companions, that can be recognized by an extra frequency modulation of the \ac{DWD} \ac{GW} signals \cite{Robson:2018,Tamanini:2018,Steffen:2018}. \ac{GW} detectors have the potential to guide the discovery of these populations \cite{dan:2014}.

TianQin is a space-based \ac{GW} observatory sensitive to millihertz frequencies \cite{Luo:2015,Hu:2019,Ye2019Optimizing}.
Recently, a significant effort has been put into the study and consolidation of the science cases for TianQin \cite{Hu:2017}. 
On the astrophysics side, these efforts include studies on the detection prospect of massive black hole binaries \cite{Wang:2019,Hu:2018}, EMRIs \cite{Fan:2019}, stellar-mass black hole binaries \cite{Liu:2019}, and stochastic backgrounds \cite{Liang:2019};
on the fundamental physics side, prospects for testing of the no-hair theorem with GWs from massive black hole binaries \cite{Shi:2019} and constraints on modified gravity theories \cite{2019PhRvD.100h4024B,Xie:2020,2019PhRvD..99j4027L,2020arXiv200301441Z} have been assessed for TianQin.
In this paper, we aim to forecast the detection of Galactic \acp{DWD} with TianQin. 
Due to their low masses, the GW horizon of \acp{DWD} is limited within the Milky Way, possibly reaching nearby satellite galaxies and the Andromeda galaxy \cite{Korol:2018b,Lamberts:2018,2020arXiv200210462K,roe:2020}. 
Therefore, in this study we focus on the Galactic population only.
We concentrate on detached systems, because they are expected to be orders of magnitude more numerous than other types of binaries in the millihertz frequency regime \cite[e.g.,][]{Nelemans:2004,Nissanke:2012}.


The paper is organized as follows.
In Section \ref{sec:2}, we outline the sample of the currently known ultra-compact \acp{DWD} and AM CVn's, and we present a mock Galactic population.
In Section \ref{sec:3}, we derive analytical expressions for computing the signal-to-noise ratio and uncertainties on binary parameters for TianQin.
In Section \ref{sec:4}, we present our results on the detectability of the known \acp{DWD} and that of the mock population. We also present similar results for some mission variations and explore the improvements that could be achieved when a few detectors work as a network.
Finally, we summarize our main findings in Section \ref{sec:5}.

\section{Galactic Double White Dwarf Binaries}\label{sec:2}
The currently known electromagnetic (EM) sample amounts to $\sim$100 detached and $\sim$60 interacting (AM~CVn) \ac{DWD} systems with orbital periods $\lesssim$1\,day \cite{Brown:2017,Maoz:2018,Ramsay:2018}.
Although rapidly expanding with several recent detections \citep{Burdge:2019a,Burdge:2019b,Coughlin:2020,Brown:2020}, this sample is still limited and represents only the tip of the iceberg of the overall Galactic population.
To quantify the ability of TianQin in detecting DWDs, in this study we consider both the known sample and a synthetic Galactic population. 
In this section, we briefly outline both samples.


\subsection{Candidate verification binaries} \label{sec:VB}

\begin{figure*}
	\begin{center}
		\includegraphics[width=1\textwidth]{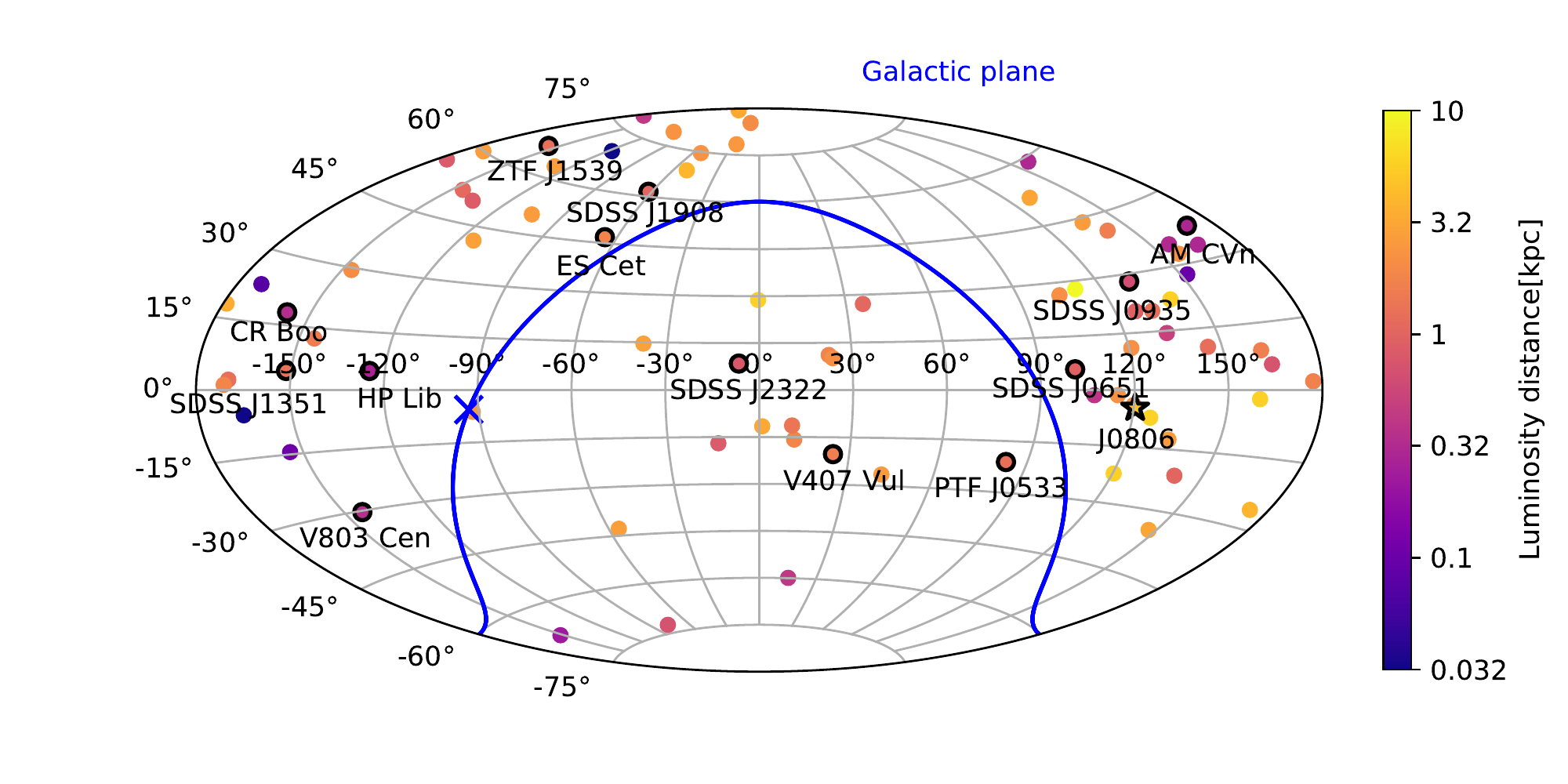}
		\caption{Sky positions of the 81 candidate \vb~shown in the ecliptic coordinate system, with lighter colors representing shorter distances to the Solar System. Binaries with the highest SNR are highlighted. The blue line indicates the Galactic plane, with the Galactic Center marked by the blue cross.}
		\label{fig:VB_sky_position}
	\end{center}
\end{figure*}

Binaries discovered through EM observations are often called \emph{verification binaries} in the literature \cite[e.g.][]{Stroeer:2006,Kupfer:2018}. This is because we can measure their parameters and therefore accurately model their \ac{GW} signals; the predicted signal can be used to verify the detector's performance.
Here we consider a sample of 81 \ac{CVBs} (40 AM~CVn type systems and 41 detached \acp{DWD}) with orbital periods $\lesssim$5\,hours. 
FIG.~\ref{fig:VB_sky_position} shows the sky positions and the luminosity distances of our \ac{CVBs} in the ecliptic coordinate system.

We list parameters of verification binaries in Table~\ref{tb:VB_parameter} in Appendix \ref{app:table-78source}.
Parameters with poor observational constraints have been inferred from theoretical models.
For example, for most verification binaries, trigonometric parallaxes from Gaia Data Release 2 \citep{gaia18} can be used to determine their luminosity distance \cite{Kupfer:2018}. 
Distances to RX J0806.3+1527 (also known as HM Cancri, hereafter J0806 \cite{2005ApJ...627..920S}), CR Boo, V803 Cen, SDSS J093506.92+441107.0, SDSS J075552.40+490627.9, SDSS J002207.65--101423.5 and SDSS J110815.50+151246.6, however, are determined using different methods.
In particular, J0806 has a largely uncertain distance.
Here we use a conservative upper boundary of 5\,kpc based on its luminosity observation \cite{Roelofs:2010}.

In this work we define a \ac{DWD} system as a verification binary if (1) it has been detected in the \ac{EM} bands, and (2) its expected \ac{GW} \ac{SNR} for TianQin is $\ge$ 5 with a nominal mission lifetime of five years \cite{Stroeer:2006,Kupfer:2018}.
We adopt a relatively low \ac{SNR} threshold for the detection of the \vb, because there is \emph{a priori} information from the \ac{EM} observations to fall back on.
We also define the potential \vb~to be the \ac{CVBs} which have $3 \leq$ SNR $< 5$ \cite{Stroeer:2006,Kupfer:2018}.

\subsection{Synthetic Galactic population} \label{sec:GB}

In this study, we employ a synthetic catalog of Galactic \acp{DWD} based on models of \citet{Toonen:2012,Toonen:2017}. These models are constructed on a statistically significant number of progenitor zero-age main sequence systems $(\sim10^5)$ evolved with binary population synthesis code {\sc SeBa} \cite{Portegies:1996} until both stars become white dwarfs.
To construct the progenitor population the mass of the primary star is drawn from the Kroupa initial mass function in the range between 0.95 and 10$\,$M$_{\odot}$ \cite{Kroupa:1993}.
Then, the mass of the secondary is drawn from a uniform mass ratio distribution between 0 and 1 \cite{Kraus:2013}.
Orbital separations and eccentricities are obtained from a the log-flat distribution (considering those binaries that on the zero-age main sequence have orbital separations up to $10^6\,R_\odot$.) 
and a thermal distribution, respectively \cite{Abt:1983,Heggie:1975,Kraus:2013}.
The binary fraction is set to 50\% and the metallicity to solar.
It is important to note that in this paper we use models that employ the $\alpha\gamma-$common envelope evolution model designed and fine-tuned on observed \acp{DWD} \cite{Nelemans:2000,Nelemans:2001a}.
We highlight that this model matches well the mass ratio distribution \citep{Toonen:2012} and the number density \citep{Toonen:2017} of the 
observed DWDs.

Next, we assign the spatial and the age distributions to synthetic binaries.
Specifically, we use a smooth Milky Way potential consistent of an exponential stellar disc and a spherical central bulge, adopting scale parameters as in \cite[][see table 1 of that source]{Korol:2018}.
The stellar density distribution is normalized according to the star formation history numerically computed by \citet{Boissier:1999}, while the age of the Galaxy is set to $13.5\,$Gyr.
We account for the change in binary orbital periods due to \ac{GW} radiation from the moment of \ac{DWD} formation until $13.5\,$Gyr.

Finally, for each binary we assign an inclination angle $\iota$, drawn randomly from a uniform distribution in $\cos \iota$.
The polarization angle and the initial orbital phase ($\psi_S$ and $\phi_0$, respectively) are randomized, assuming uniform distribution over the intervals of [0, $\pi$) and [0, 2$\pi$), respectively.
The obtained catalog contains the following parameters:
orbital period $P$, component masses $m_1$ and $m_2$,
the ecliptic latitude $\lambda$ and longitude $\beta$, distance from the Sun $d$, and angles $\iota, \psi_S, \phi_0$.
This catalog has been originally employed in the study of \ac{DWD} detectability with LISA \citep{LISA:2017}. Therefore, this paper represents a fair comparison with the results in \citet{LISA:2017}.

\section{Signal and noise modeling}\label{sec:3}
\subsection{Gravitational wave signals from a monochromatic source} \label{sec:3a}

The timescale on which \acp{DWD}' orbits shrink via \ac{GW} radiation is typically $>$ Myr (at low frequencies). 
This is significantly greater than the mission lifetime of TianQin of several years; the two timescales are only comparable when $f/\dot{f} \sim T_\mathrm{m}$:
\be f = 0.18 \left( \frac{T_{\mathrm{m}}}{5 \mathrm{yr}} \right)^{-3/8} \left( \frac{\mathcal{M} }{1 M_\odot} \right)^{-5/8} \mathrm{Hz} 
\label{eq:monochromatic_codition} \ee
Therefore, binaries with frequencies significantly smaller than 0.18 Hz can be safely considered as monochromatic \ac{GW} sources, meaning  that they can be described by a set of seven parameters:
the dimensionless amplitude $(\mathcal{A})$, \ac{GW} frequency $f=2/P$, $\lambda$, $\beta$, $\iota$, $\psi_S$ and $\phi_0$.
Note that we do not include eccentricity because \acp{DWD} circularize during the common envelope phase. 

GWs emitted by a monochromatic source can be computed using the quadrupole approximation \cite{Landau:1962,Peters:1963}.
In this approximation, the \ac{GW} signal can be described  as a combination of the two polarizations ($+$ and $\times$): 
\be h_+(t) = \mathcal{A}(1+\cos\iota^2)\cos(2\pi ft + \phi_0 + \Phi_D(t))\,,
\label{eq:hplus} \ee
\be h_{\times}(t) = 2\mathcal{A}\cos\iota\sin(2\pi ft + \phi_0 + \Phi_D(t))\,,
\label{eq:hcross} \ee
with
\be \label{eq:GWamp}
\mathcal{A} = \frac{2(G\mathcal{M})^{5/3}}{c^4d}(\pi f)^{2/3}, \ee 
where $\mathcal{M}\equiv (m_1 m_2)^{3/5}/(m_1+m_2)^{1/5}$ is the chirp mass, and $G$ and $c$ are the gravitational constant and the speed of light, respectively.
Note that the additional term $\Phi_D(t)$ in the \ac{GW} phase [Eq.~\eqref{eq:hplus}-\eqref{eq:hcross}] is the Doppler phase arising from the periodic motion of TianQin around the Sun:
\be \Phi_D(t) = 2\pi f t \frac{R}{c} \sin(\pi/2-\beta) \cos(2\pi f_m t-\lambda)\, ,
\label{} \ee
where $R=1\mathrm{A.U.}$ is the distance between the Earth and the Sun, and $f_m=1/\mathrm{year}$ is the modulation frequency.
$\lambda$ and $\beta$ are the ecliptic coordinates of the source. 

\subsection{Detector's response to \ac{GW} signals}

The design of the TianQin mission \cite{Luo:2015} envisions a constellation of three drag-free satellites orbiting the Earth, maintaining a distance between each other of $\sim10^5$\,km. 
Satellites will form an equilateral triangle constellation oriented in such a way that the normal vector to the detector's plane is pointing towards J0806 ($\lambda=120.4^\circ$, $\beta=-4.7^\circ$). 

\renewcommand\arraystretch{1.5}   
\begin{table*}
\begin{tabular}{l | c }
\hline
Configuration          &TianQin    \\
\hline
Number of satellites   & N=3 \\
Orientation            & $\lambda=120.4^\circ$, $\beta=-4.7^\circ$  \\
Observation windows    &2 $\times$ 3 months each year  \\
Mission lifetime       & 5 years   \\
Arm length             & $L = \sqrt{3}\times10^5\mathrm{km}$   \\
Displacement measurement noise         & $S_x = 1\times10^{-24}\mathrm{m}^2\mathrm{Hz}^{-1}$  \\
Acceleration noise     & $S_a = 1\times10^{-30}\mathrm{m}^2\mathrm{s}^{-4}\mathrm{Hz}^{-1}$  \\
\hline
\end{tabular}
\caption{Key parameters for the TianQin configurations.}
\label{tb:configuration}
\end{table*}


In the low-frequency limit ($f\ll f_*$ with $f_*=c/2\pi L$ being the transfer frequency,  $\sim0.28$ Hz for TianQin), the \ac{GW} strain recorded by the detector can be described as a linear combination of the two \ac{GW} polarizations modulated by the detector's response \cite{Cutler:1998}:
\be h(t) =h_+(t)F^+(t)+h_{\times}(t)F^{\times}(t)\,,
\label{eq:waveform_1} \ee
where $F^{+,\times}(t)$ are the antenna pattern functions.

For a detector with an equilateral triangle geometry, two orthogonal Michelson signals can be constructed and
the antenna pattern functions can be expressed as 
\begin{widetext}
\begin{eqnarray}
F^+_{\mathrm{1}}(t,\theta_S,\phi_S,\psi_S) &=& \frac{\sqrt{3}}{2}\Bigg(\frac{1}{2}(1+\cos^2\theta_S)\cos2\phi_S(t)\cos2\psi_S  \label{eq:Fplus}   -\cos\theta_S\sin2\phi_S(t)\sin2\psi_S \Bigg) \, , \\
F^\times_{\mathrm{1}}(t,\theta_S,\phi_S,\psi_S) &=& \frac{\sqrt{3}}{2}\Bigg(\frac{1}{2}(1+\cos^2\theta_S)\cos2\phi_S(t)\sin2\psi_S   +\cos\theta_S\sin2\phi_S(t)\cos2\psi_S\Bigg) \, ,\\
F^+_{\mathrm{2}}(t,\theta_S,\phi_S,\psi_S)&=&F^+_{\mathrm{1}}(t,\theta_S,\phi_S -\frac{\pi}{4},\psi_S) \, ,\\
F^\times_{\mathrm{2}}(t,\theta_S,\phi_S,\psi_S) &=&F^\times_{\mathrm{1}}(t,\theta_S,\phi_S-\frac{\pi}{4},\psi_S)\, , \label{eq:Fcross}
\end{eqnarray}
\end{widetext}
where $\sqrt{3}/2$ represents a factor originating from the geometry of the detector and encodes the $60^{\circ}$ angle between the detector's arms, and $\theta_S$ and $\phi_S(t)=\phi_{S0}+\omega t$ are the latitude and longitude of the source in the detector's coordinate frame, with $\omega\approx 2 \times 10^{-5}$ rad/s being the angular frequency of the TianQin satellites.
The transformation from the ecliptic coordinates ($\beta,\lambda$) to the detector coordinates ($\theta_S,\phi_S$) can be found in Appendix \ref{Coordinate transformation}.
The subscripts 1 and 2 in Eqs.~(\ref{eq:Fplus})-(\ref{eq:Fcross}) are labels for the two Michelson signals, which are orthogonal to each other, as indicated by the $\pi/4$ phase difference between the corresponding antenna pattern functions \cite[e.g.,][]{Cutler:1998}.
From Eqs.~(\ref{eq:Fplus})-(\ref{eq:Fcross}) one can conclude that TianQin is most sensitive to GWs propagating along the normal direction to the detector's plane, and least sensitive to \acp{GW} propagating along the detector plane.

In general, the exact inclusion of the antenna pattern functions is complicated \cite[e.g.,][]{Cornish:2003}.
In practice, we introduce the sky-averaged response function $R(f)$ to simplify the following calculations.
It can be approximated by
\begin{eqnarray}
R(f)\approx\frac{3}{10}\frac{1}{1+0.6(f/f_*)^2}\,.
\label{eq:heffbar}
\end{eqnarray}
The prefactor $3/10=2\times3/20$ is two times (to account for two independent Michelson interferometers) the sky-averaged factor of $3/20$, which can be obtained as $\overline{F^{\times,+}}$, with $\overline{F^{\times,+}}\equiv\frac{1}{4\pi^2}\int^\pi_0 \mathrm{d}\psi_S \int^{2\pi}_0 \mathrm{d}\phi \int^\pi_0 F^{\times,+}\sin\theta \mathrm{d}\theta$.

\subsection{Detector noise and the scaled sensitivity curve} \label{sec:sensitivity}

\begin{figure*}
\begin{center}
\includegraphics[width=0.5\textwidth]{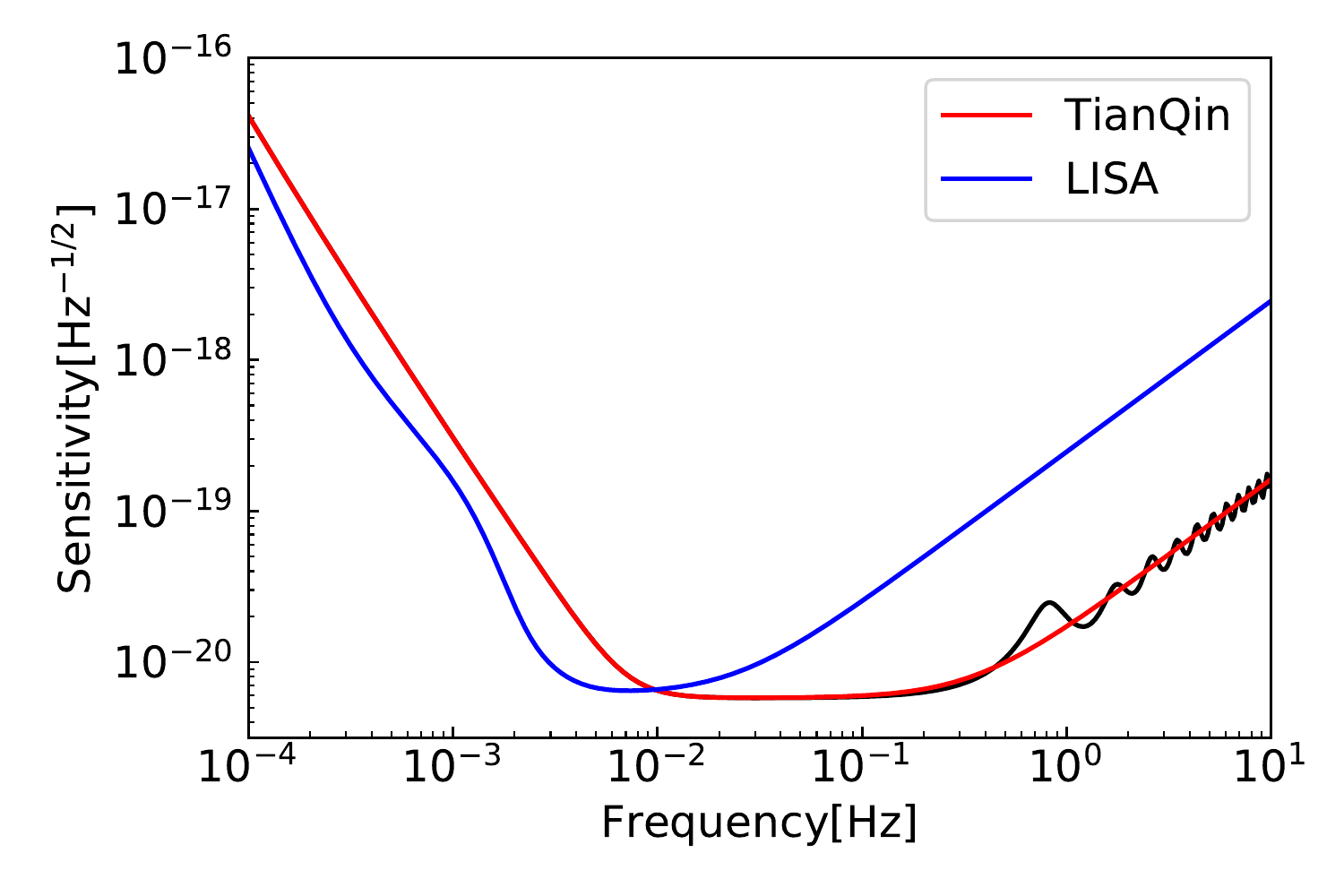}
\caption{The sensitivity curve of TianQin. The red line corresponds to $\td{S}_n(f)$, defined in Eq.~(\ref{eq:S_n^NSA}), while the black line corresponds to the full sky-averaged result, preserving all the frequency dependence (see Eqs.~(15)-(16) in \cite{Wang:2019}). 
The LISA sensitivity curve is shown by the blue line according to \cite{Robson:2019}.}
\label{fig:TianQin_ASD}
\end{center}
\end{figure*}

The huge number of Galactic \acp{DWD} can generate a foreground confusion noise that may affect the detection of other types of \ac{GW} sources. In Sect.~\ref{sec:foreground} we show that such foreground is relatively weak for TianQin. 
Therefore, through this paper we only consider the instrumental noise.

The noise spectral density  of TianQin can be expressed analytically as
\bea S_N(f) &=& \frac1{L^2}\left[\frac{4S_a}{(2\pi f)^4}\left(1+\frac{10^{-4}{\rm Hz}}{f}\right) +S_x\right],
\label{eq:S_N} \eea
where $L$, $S_a$, $S_x$ are given in Table \ref{tb:configuration}.

From the sky-averaged response function Eq.~(\ref{eq:heffbar}) and the detector noise Eq.~(\ref{eq:S_N}), one can construct the sensitivity curve of the detector as
\bea \td{S}_n(f) &=&S_N(f)\Big/\td{R}(f)\nn\\
&=&\frac{1}{L^2}\left[\frac{4S_a}{(2\pi f)^4}\left(1+\frac{10^{-4} Hz}{f}\right) +S_x\right]\nn\\ &&\times\left[ 1+0.6\left(\frac{f}{f_*}\right)^2 \right]\,,
\label{eq:S_n^NSA} \eea
where
\begin{eqnarray}
  \tilde{R}(f)\equiv R(f)\frac{10}{3}=\frac{1}{1+0.6(f/f_*)^2}\,.
\label{eq:Rtd}
\end{eqnarray}
Note that in this formalism, we assume the effect of the antenna pattern to be associated with the signal.
The obtained sensitivity curve is represented in FIG.~\ref{fig:TianQin_ASD}.
\textbf{}

\subsection{Data analysis} \label{sec:snr}

The \ac{SNR} $\rho$ of a signal is defined as
\be  \rho^2 = (h|h),
\label{eq:SNR}\ee
where the inner product $(\cdot|\cdot)$ is defined as \cite{Finn:1992,Cutler:1994},
\bea (a|b)&=&4\Re e\int^\infty_0 \mathrm{d}f \frac{\tilde{a}^*(f)\tilde{b}(f)}{\td{S}_n(f)}\nn\\
&\simeq&\frac{2}{\td{S}_n(f_0)}\int^{T}_0 \mathrm{d}t \,a(t)b(t),
\label{eq:inner product} \eea
where $\tilde{a}(f)$ and $\tilde{b}(f)$ are the Fourier transformations of two generic functions $a(t)$ and $b(t)$, $\td{S}_n(f)$ is defined in Eq.~\eqref{eq:S_n^NSA}.
The second step is obtained by using Parseval's theorem and the quasi-monochromatic nature of the signal, which acts like a Dirac delta function on the noise power spectral density \cite{Cutler:1998}.

For a monochromatic \ac{GW} signal with frequency $f_0$, it is possible to derive an analytical expression of the SNR ($\rho$)\footnote{Note that our definitions of the SNR and the amplitude differ from those in \cite{Korol:2017} by a numerical factor, but we both are self-consistent.}

\be  \rho^2 = (h|h) \simeq \frac{2}{\td{S}_n(f_0)}\int^{T}_0 \mathrm{d}t \,h(t)h(t) =\frac{2\langle A^2\rangle T}{\td{S}_n(f_0)}\,,
\label{eq:SNR_mono} \ee
with
\begin{eqnarray}
\langle A^2\rangle & = & \frac{1}{T}\int^T_0 h^2(t) \,\mathrm{d}t.\label{eq:def_ave_amp}\\
&\approx& \frac{3}{16}\mathcal{A}^2 \Big[ (1+\cos^2\iota)^2 \langle  F_{+}^2\rangle\nn\\
&&\qquad\quad+4\cos^2\iota\langle F_{\times}^2\rangle\Big] ,   \label{eq:ave_amp}\\
\langle F_{+}^2\rangle &=&\frac{1}{4}(1+\cos^2\theta_S)^2\cos^22\psi_S\nn\\
&&+\cos^2\theta_S\sin^22\psi_S , \label{eq:ave_plus}\\
\langle F_{\times}^2\rangle &=& \frac{1}{4}(1+\cos^2\theta_S)^2\sin^22\psi_S\nn\\
&&+\cos^2\theta_S\cos^22\psi_S\,, \label{eq:ave_f}
\end{eqnarray}
where $T$ is the observation time (which is half the operation time), and we have neglected the $\mathcal{O}(T^{-1})$ terms in Eq.~(\ref{eq:ave_amp}).
It is also useful to define the characteristic strain $h_c = A \sqrt{N}$, with $N=f_0T$ being the number of binary orbital cycles observed during the mission.
Analogously, the noise characteristic strain is $h_n(f)=\sqrt{f\td{S}_n(f)}$.
One can straightforwardly estimate the \ac{SNR} from the ratio between $h_c$ and $h_n$.

\subsection{Galactic \ac{GW} foreground}\label{sec:foreground_result}

At frequencies $<1$\,mHz, the number of Galactic sources per frequency is too large to resolve all individual \ac{GW} signals. These signals can potentially become indistinguishable and form a foreground for the TianQin mission \cite[in analogy with][]{Cornish:2017vip}.
We assess the level of such a foreground using a synthetic population presented in Section \ref{sec:GB}.

We follow the method outlined in \citet{Littenberg:2014}.
For each binary, we construct the signal in the frequency domain, $h(f)$ (cf. Section \ref{sec:3a}).
All signals in each frequency bin are then incoherently added, forming an overall population spectrum.
Next, we smooth the spectrum by a running median smoothing function with a set window size and by fitting with cubic spline to it. We define the smoothed Galactic spectrum $S_{\rm DWD}(f)$ and compute the total noise as the sum of the instrumental noise $S_n(f)$ and $S_{\rm DWD}(f)$.
Using the updated noise curve, we check if any \acp{DWD} results have a \ac{SNR} larger than the preset threshold of 7. These ``resolved" \acp{DWD} are then removed from the sample, and the process is repeated from the beginning.
The iterations are performed until the convergence i.e., until there are no more new resolved sources. The final result is represented in FIG.~\ref{fig:Foreground}.

\begin{table}[]
	\begin{center}
		\caption{The coefficients for the polynomial fit for the foreground, as $10^{\sum_i a_{i}x^{i}}$, where $x=\log{(f/10^{-3})}$. Successive rows correspond to increasing operation time $T$.}
	    \centering
	    \label{tb:DWD}
	    \begin{tabular}{c c c c c c c c}
		    \hline
		    $T$   &$a_{0}$ & $a_{1}$&$a_{2}$ &$a_{3}$& $a_{4}$&$a_{5}$ &$a_{6}$\\
		    \hline
		    \hline
	    	0.5 yr  & -18.6  & -1.22   & 0.009 &  -1.87  &  0.65  &  3.6  & -4.6\\
	    	\hline
	    	1 yr    & -18.6  & -1.13   & -0.945  &  -1.02  &  4.05   &  -4.5 & -0.5\\
	    	\hline
	    	2 yr    & -18.6  & -1.45   & 0.315   &  -1.19  & -4.48   &  10.8  & -9.4 \\
	    	\hline
	    	4 yr    & -18.6  & -1.43   & -0.687  &  0.24  & -0.15  & -1.8  & -3.2\\
    		\hline
	    	5 yr    & -18.6  & -1.51   & -0.710  & -1.13   & -0.83  &  13.2  & -19.1\\
	    	\hline
    	\end{tabular}
    \end{center}
\end{table}

\subsection{Parameter estimation}
The uncertainty on the binary parameters can be derived from the \ac{FIM} $\Gamma_{ij}$,
\be \Gamma_{ij} = \left(\frac{\partial h}{\partial \xi_i}\bigg|\frac{\partial h}{\partial \xi_j}\right),
\label{eq:FIM} \ee
where $\xi_{i}$ stands for the $i{\rm th}$ parameter.

In the high-\ac{SNR} limit ($\rho \gg$ 1), the inverse of the FIM equals to the variance-covariance matrix, $\Sigma = \Gamma^{-1}$.
The diagonal entries $\Sigma_{ii}$ give the variances (or mean square errors) of each parameter, $(\Delta \xi_i)^2 $, while the off-diagonal entries describe the covariances.
In numerical calculations, we approximate $\partial h /\partial \xi_i$ with numerical differentiation
\begin{eqnarray}\label{eq:delta}
\frac{\partial h}{\partial\xi_i} \approx  \frac{\delta h}{\delta\xi_i}\equiv \frac{h(t,\xi_i+\delta\xi_i)-h(t,\xi_i-\delta\xi_i)}{2\delta\xi_i}\,.
\end{eqnarray}\\
The differentiation steps $\delta\xi_i$ were chosen to make the numerical calculation stable \cite{Shah:2012}.

Notice that compared with the uncertainty of each coordinate, we are more interested in the sky localization, which is a combination of the uncertainties of both coordinates \cite{Cutler:1998}:
\be \Delta\Omega_{\rm S} = 2\pi \big|\sin\beta \big| \big(\Sigma_{\beta\beta}\Sigma_{\lambda\lambda}-\Sigma^2_{\beta\lambda} \big)^{1/2}\,.
\label{eq:Omega} \ee

When a network of independent detectors is considered, the total \ac{SNR} and \ac{FIM} of a source can be calculated as
\bea  \rho^2_{\mathrm{total}}&=& \sum_{a} \rho^2_a = \sum_{a} (h_a|h_a)\,,\nn\\
\Gamma_{\mathrm{total}}&=&\sum_{a} \Gamma_a = \sum_{a} \left(\frac{\partial h_a}{\partial \xi_i}\bigg|\frac{\partial h_a}{\partial \xi_j}\right)\,,
\label{eq:multi} \eea
where the subscript $a$ stands for quantities related to the $a{\rm th}$ detector.

\section{Results}\label{sec:4}
In this section we report our results for the TianQin mission. We also consider an alternative version of the mission configuration with the same characteristics (cf. Table~\ref{tb:configuration}), but oriented perpendicularly to the original TianQin's configuration (pointing towards $\lambda=30.4^\circ$ and $\beta=0^\circ$). 
In the following, we denote the standard TianQin configuration as TQ and the additional one as TQ II.
\ac{GW} observations can be improved if many detectors are working simultaneously in a network (e.g., the LIGO + Virgo network). 
Therefore, in this work we also explore the possibility of two detectors TQ and TQ II operating simultaneously, both following the ``three months on + three months off" observation scheme as a way to fill the data gaps of each other. 
We refer to the configuration consisting of the two detectors as TQ I+II. 
In addition, we also explore the possibility of TQ and TQ I+II operating together with LISA.

\subsection{Galactic foreground}\label{sec:foreground}

\begin{figure*}
	\centering
	\includegraphics[width=0.6\textwidth]{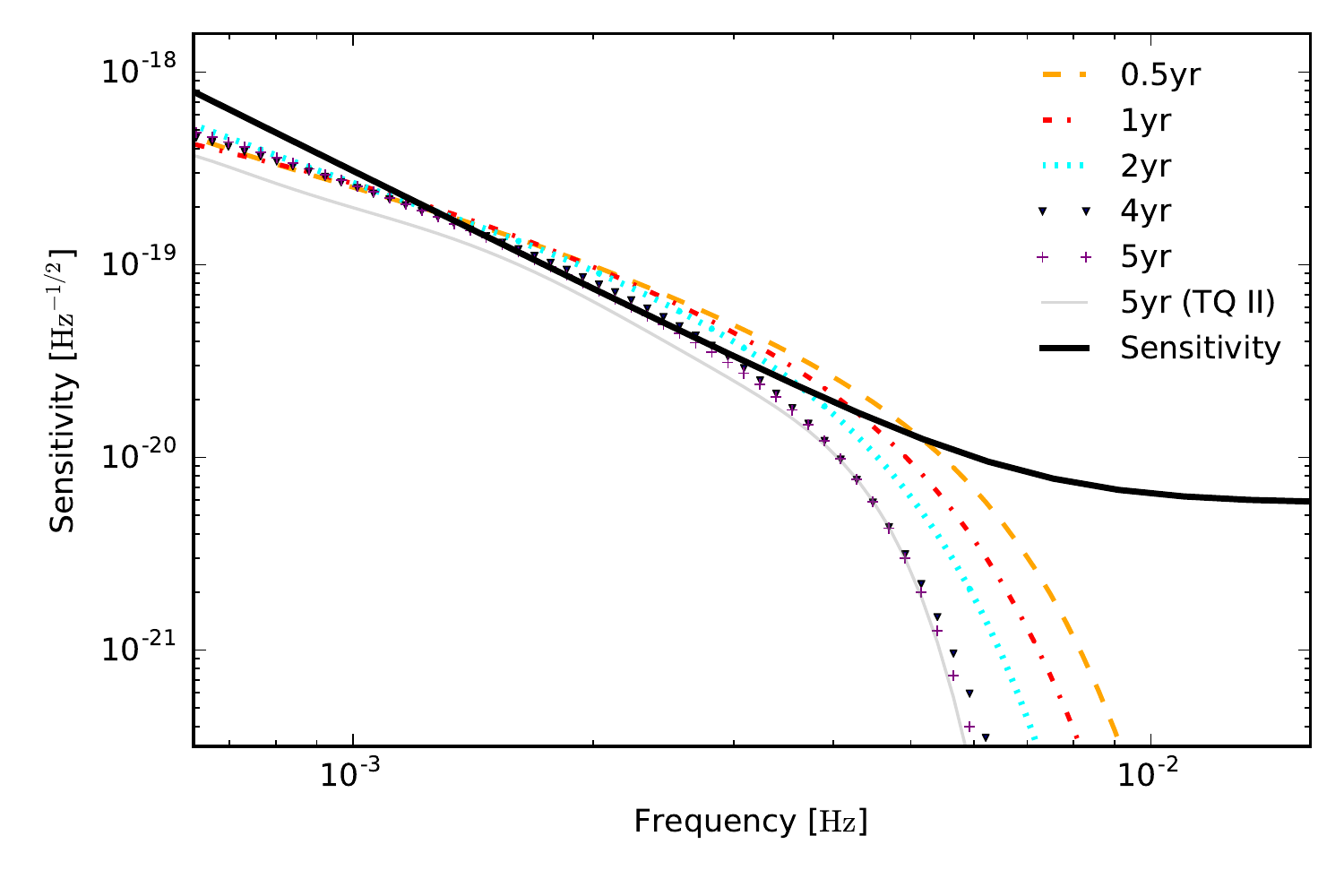}
	\caption{Expected foreground from Galactic DWDs for different operation times (colored lines). The black solid line represents the sensitivity curve of TianQin (cf. FIG.~\ref{fig:TianQin_ASD}).
        The foreground with a 5 year operation time for TianQin II is also shown for comparison. Notice that it is consistent with the foreground of the TianQin constellation with the same operation time.
		}
	\label{fig:Foreground}
\end{figure*}

First, we assess the impact of the Galactic confusion foreground for TianQin. In FIG.~\ref{fig:Foreground}, we show the estimates of the foreground levels corresponding to different operation times (colored lines) obtained according to the procedure described in Section~\ref{sec:foreground}. 
Each line can be reproduced by using the expression $S_{\rm DWD}(f)=10^{\sum_i a_{i}x^{i}}$,
where $x=\log{(f/10^{-3})}$ and polynomial coefficients $a_i$ are reported in Table \ref{tb:DWD} for different operation times.

From FIG.~\ref{fig:Foreground}, it is evident that the foreground strain is inversely proportional to the operation time.
Therefore we did not include the Galactic foreground in the following analysis.
Notice that the foreground of TianQin and TianQin II are quite consistent, so change in orientation has relatively minor effect on the overall foreground.
This is illustrated in FIG.~\ref{fig:Foreground} where foreground of TQ II for 5 year operation time is shown.

\subsection{Verification binaries}

\begin{figure*}
	\includegraphics[width=1\textwidth]{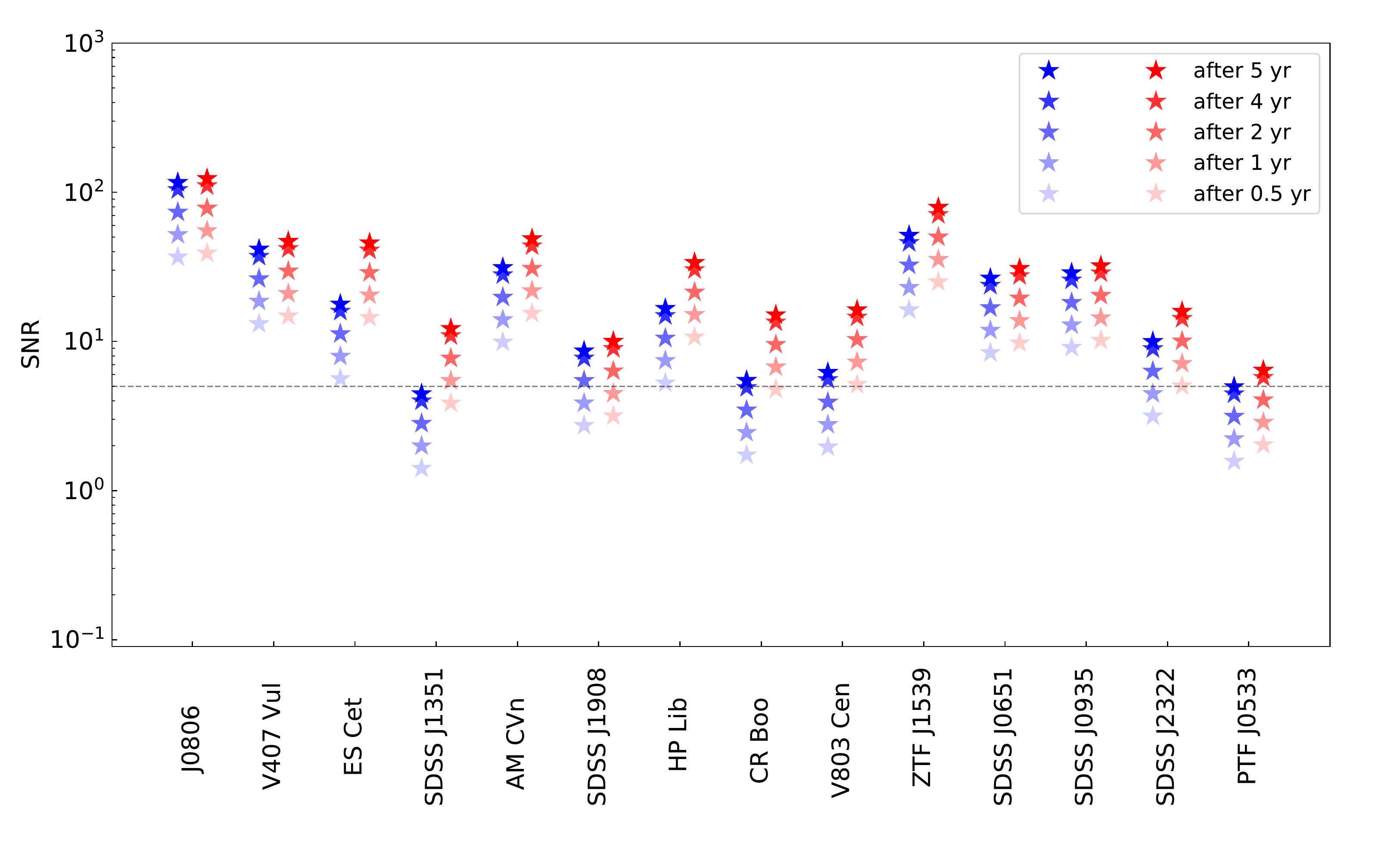}
	\caption{The \ac{SNR} evolution of \vb~over time.
		Blue stars represent TianQin (TQ), and red stars represent TQ I+II.
		The black dashed line corresponds to the SNR threshold of 5.}\label{fig:VB_SNR_evolution}
\end{figure*}

Out of 81 considered candidates (cf. Table~\ref{tb:VB_parameter}) we find 12 \vb~with SNR$\ge5$: J0806, V407 Vul, ES Cet, AM CVn, SDSS J1908, HP Lib, CR Boo, V803 Cen, ZTF J1539, SDSS J0651, SDSS J0935, and SDSS J2322, with J0806 having the highest SNR.
In particular, we find that J0806 reaches a SNR threshold of 5 already after only two days of observation. We predict that its SNR will reach 36.8 after three months of observation, and will exceed 100 after nominal five years of mission (effectively corresponding to 2.5 years of observation time).
In addition, we find three potential \vb~with 3\,$\le$\,SNR\,$<$\,5: SDSS J1351, CXOGBS J1751 and PTF J0533.
Figure \ref{fig:VB_SNR_evolution} shows the evolution of the SNR with time for all \vb~in blue for TianQin (TQ).

In Table~\ref{tb:VB_SNR}, we report dimensionless amplitudes ($\mathcal{A}$) and SNRs for all 81 \ac{CVBs} considering mission configurations: TianQin (TQ), TQ II, and TQ I+II, assuming five years of mission lifetime and setting $\phi_0=\pi$ and $\psi_S=\pi/2$ for all binaries.
We note that the sky position, orbital inclination and \ac{GW} frequency of the binary affect SNR by a factor of a few (cf. Eqs.\eqref{eq:SNR_mono}-\eqref{eq:ave_f}).
For example, V803 Cen and SDSS J0651 have comparable \ac{GW} amplitudes ($16.0\times10^{-23}$ and $16.2\times10^{-23}$, respectively), but their SNRs differ significantly (6.2 and 26.5, respectively). 
This difference arises both from the fact that SDSS J0651 is located in a more favorable position in the sky for TianQin (TQ), and the fact that it has a higher frequency than V803 Cen.
We also note that because TianQin (TQ) is oriented directly towards J0806, its SNR is the largest across the sample, although its amplitude is not the highest. When considering the TQ II configuration with a different orientation, its SNR decreases by a factor of $\sim3$.

We find that TQ II can detect 13 \vb~with SNR$>5$ and one potential verification binary with 3\,$<$\,SNR\,$<$\,5. 
Being orthogonal to TianQin (TQ), the TQ II configuration is more disadvantageous than for the detection of J0806.
However, even with TQ II, J0806 can be detected with a SNR of 41.6. 
This is because J0806 has the highest frequency across the sample (cf. Table~\ref{tb:VB_parameter}). 
With a frequency of 6.22\,mHz, it is positioned in the amplitude-frequency parameter space where the noise level of TianQin is the lowest (see also FIG.~\ref{fig:VB_GB_ASD}).
Therefore, J0806 is still among the best verification sources for the TQ II configuration.

Similarly, for the network TQ I+II there will be 14 \vb~and 1 potential verification binary. 
The SNR evolution for different operation times for these \vb~is represented in red in FIG.~\ref{fig:VB_SNR_evolution}. 
The SNR produced by a source in this case is given by the root sum squared of the SNRs of the two configurations considered independently (see Eq.\eqref{eq:multi}).
Therefore, if TianQin (TQ) and TQ II independently detect a source with a similar SNR, the network TQ I+II would improve the SNR by a factor of $\sqrt{2}$.
However, if the source produces significantly higher SNR in one of the detectors in the network, the improvement is not significant (e.g., J0806 in Table~\ref{tb:VB_SNR}).

In Table \ref{tb:FIM}, we fix all other parameters and only report the estimated uncertainties on the amplitude $\mathcal{A}$ and inclination angle $\iota$ for the 14 \vb. 
These two parameters are typically degenerate (cf. Eqs.~\eqref{eq:hplus}-\eqref{eq:hcross}).
However, for nearly edge-on binaries the degeneracy can be broken by using the asymmetry between two \ac{GW} polarizations \citep[e.g.][]{Shah:2012}. 
We present TianQin's ability to constrain the polarization angle $\psi$ in later sections. We remark that \cite{Xie:2020} demonstrate that in the most favorable setup, TianQin has the potential to constrain the relative strength of extra polarization modes to the tensor modes at an accuracy of about $10^{-3}$.
This is reflected in a small correlation coefficient $c_{\mathcal{A},\cos\iota}=0.157$ of SDSS J0651 with the inclination angle of $86.95^{\circ}$.
For decreasing inclination angles the degeneracy increases as can be seen for 
SDSS J1908 and V803 Cen with inclination angles of $\iota=15^{\circ}$ 
and $\iota=13.5^{\circ}$ respectively. 
These two \vb~have $\Delta{\cos\iota}>1$, meaning that the uncertainty on the inclination angle exceeds the physical range (0, $\pi$).

By means of like eclipsing binaries observations, the inclination angle can be independently determined from the \ac{EM} channel.
It can then be used to narrow down the uncertainty on the inclination from \ac{GW} data by removing the respective row and column of the FIM. 
In the column denoted ``With EM on $\iota$'' in Table \ref{tb:FIM}, we recalculate the uncertainties on the amplitude by inverting $\Gamma_{\mathcal{A}\mathcal{A}}$, equivalently assuming that the inclination of the binary is known by EM observation, and we report the ratio between this uncertainty and the uncertainty estimated without EM observation on $\iota$ (fourth column of Table \ref{tb:FIM}). 
We find that, when the inclination angle is known {\it a priori}, the uncertainty on the amplitude can be improved up by to a factor of $\sim$ 16 (e.g., for SDSS J2322), depending on the exact value of the inclination angle of the source.
Note that the improvement for nearly edge-on binaries (ZTF J1539 and SDSS J0651) is negligible. 

\begin{table}[]
	\centering
        \caption{Uncertainties on $\mathcal{A}$ and $\iota$ for 14 \vb~considering the TQ I+II configuration. In the column denoted ``Without EM on $\iota$,'' we report uncertainties on $\mathcal{A}$ and $\cos\iota$ derived from inverting the $2\times2$ FIM. In the column denoted ``With EM on $\iota$,'' we report uncertainties on $\mathcal{A}$ for the case when $\iota$ is known {\it a priori} from EM observation.}
	
	\label{tb:FIM}
	\begin{tabular}{l|ccc|cc}
		\hline 
		\multirow{2}{*}{Source} &
		\multicolumn{3}{c|}{without EM on $\iota$} &
		\multicolumn{2}{c}{with EM on $\iota$} \\ \cline{2-6} 
		&
		\multicolumn{1}{c}{$\Delta \mathcal{A}/\mathcal{A}$} &
		\multicolumn{1}{c}{$\Delta{\cos\iota}$} &
		\multicolumn{1}{c|}{$c_{\mathcal{A},\cos\iota}$} &
		\multicolumn{1}{c}{$\Delta'{\mathcal{A}}/\mathcal{A}$} &
		\multicolumn{1}{c}{$\Delta{\mathcal{A}}/\Delta'{\mathcal{A}}$}\\ \hline \hline
		J0806      & 0.061& 0.055& 0.991& 0.008& 7.625\\
		V407 Vul   & 0.050& 0.039& 0.904& 0.021& 2.381\\
		ES Cet     & 0.051& 0.039& 0.904& 0.022& 2.318\\
		SDSS J1351 & 0.193& 0.145& 0.905& 0.082& 2.354\\
		AM CVn     & 0.115& 0.102& 0.984& 0.020& 5.750\\
		SDSS J1908 & 6.102&  $>1$& 1.000& 0.100& $>1$\\
		HP Lib     & 0.384& 0.360& 0.997& 0.030& 12.800\\
		CR Boo     & 0.865& 0.813& 0.997& 0.066& 13.106\\
		V 803 Cen  & 4.377&  $>1$& 1.000& 0.062& $>1$\\
		ZTF J1539  & 0.013& 0.012& 0.300& 0.013& 1.000\\
		SDSS J0651 & 0.033& 0.018& 0.157& 0.032& 1.031\\
		SDSS J0935 & 0.073& 0.056& 0.904& 0.031& 2.355\\ 
		SDSS J2322 & 1.033& 0.979& 0.998& 0.063& 16.397\\
		PTF J0533  & 0.219& 0.137& 0.700& 0.156& 1.404\\\hline
	\end{tabular}
\end{table}

\subsection{Simulated Galactic double white dwarf binaries}

To forecast the total number of binaries detectable by TianQin we employ the simulated population of Galactic \acp{DWD} (cf. Sec.~\ref{sec:GB}). Here, we set a higher \ac{SNR} threshold of 7, assuming that there is no \emph{a priori} information from the \ac{EM} observations to fall back on.

\begin{figure*}
\includegraphics[width=0.8\textwidth]{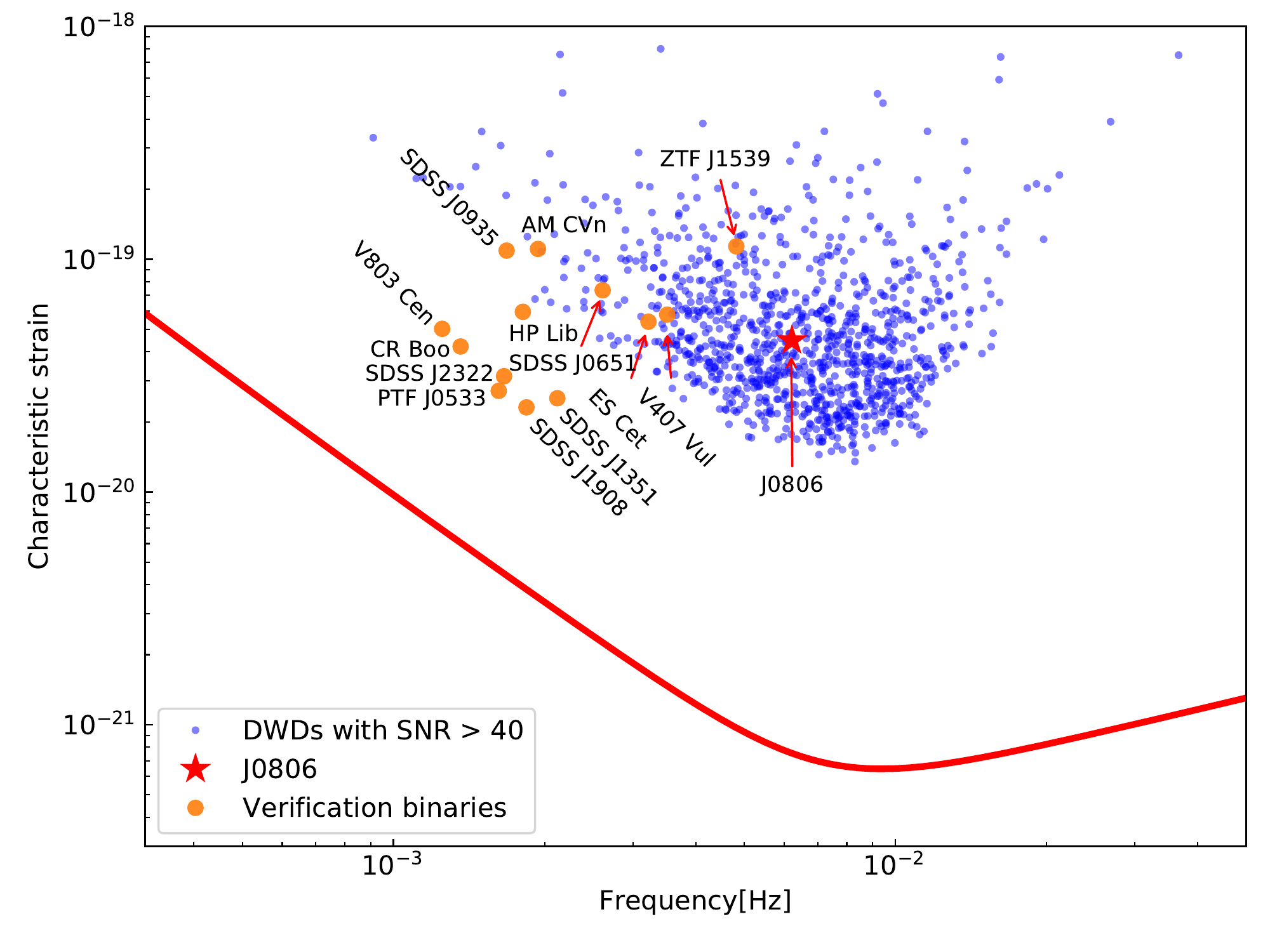}
  \caption{The characteristic strain $h_c$ of the 14 \vb~(the golden dots and the red star) for TQ I+II and the simulated \acp{DWD} with \ac{SNR} $>$ 40 for TianQin, compared with the noise amplitude $h_n$ of TianQin (red line). J0806 is highlighted with a red star. An operation time of 5 years is assumed.}
\label{fig:VB_GB_ASD}
\end{figure*}

We estimate the number of resolved \acp{DWD} for the three considered configurations (TQ, TQ II, and TQ I+II) to be of the order of several thousand for the full mission lifetime of 5 years. In Table \ref{tb:resolved_DWDs}, we summarize our result for increasing operation times. 
In FIG.~\ref{fig:VB_GB_ASD} we show the dimensionless characteristic strain of \ac{DWD}s with SNR$>40$ in the mock population compared to 14 \vb.

The density of \acp{DWD} in the bulge region of the Galaxy is significantly higher than in the disk (see Fig.~3 of \citet{Korol:2018}); therefore the detector's orientation has a significant impact on the total number of detectable \acp{DWD}.
The Galactic Center (where the density of \acp{DWD} is the highest) in ecliptic coordinates corresponds to $(\lambda=266.8^\circ, \beta=-5.6^\circ)$. 
TianQin (TQ) is oriented towards $(\lambda=120.4^\circ, \beta=-4.7^\circ$) that is, about 30$^\circ$ away from the Galactic Center; TQ II is oriented towards ($\lambda=30.4^\circ, \beta=0^\circ$),which is about 60$^\circ$ away from the Galactic Center.
Consequently, the number of detected \acp{DWD} for TianQin (TQ) is about $1.3$-$1.4$ times larger than for TQ II (cf. Table \ref{tb:resolved_DWDs}).
When we consider TQ I+II, the number of detections increases by $\sim1.3$ compared to TianQin (TQ) alone.
We verify that pointing the detector towards the Galactic Center would return the maximum detections $\sim1.0\times10^4$.

\renewcommand\arraystretch{2}   
\begin{table}
\begin{tabular}{l r r r r r}
\hline
                   &0.5yr&  1yr&  2yr&  4yr& 5yr   \\
\hline
\hline
TQ    & 2371& 3589& 5292& 7735& 8710  \\
\hline
TQ II   & 1672& 2595& 3943& 5782& 6540  \\
\hline
TQ I+II & 3146& 4716& 6966& 10023& 11212\\
\hline
\end{tabular}
  \caption{The expected detection numbers of resolvable binaries for TianQin (TQ), TQ II, and TQ I+II.} \label{tb:resolved_DWDs}
\end{table}

\begin{figure*}
\centering
\begin{minipage}{0.5\textwidth}
\includegraphics[width=0.8\textwidth]{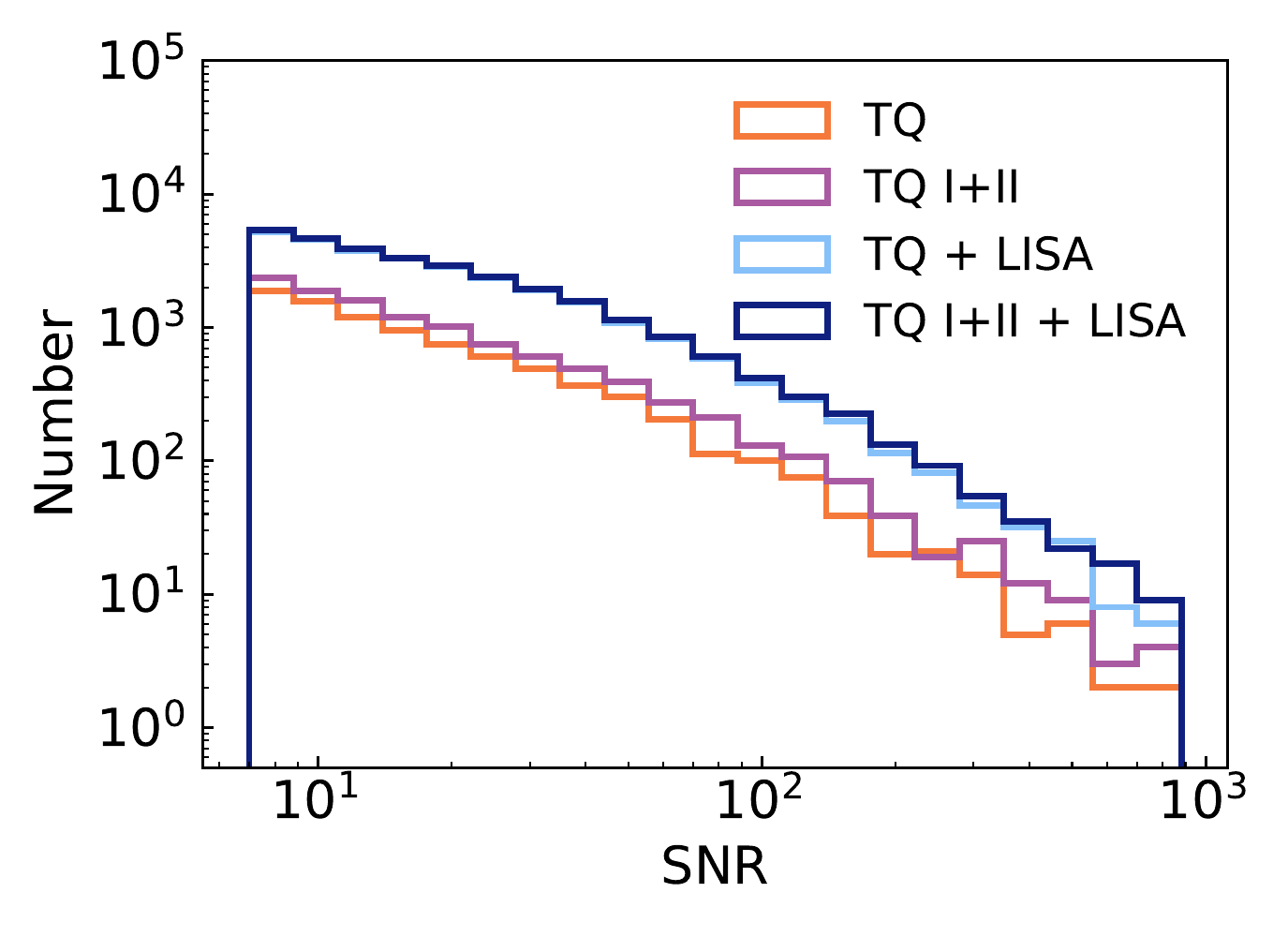}
\end{minipage}%
\begin{minipage}{0.5\textwidth}
\includegraphics[width=0.8\textwidth]{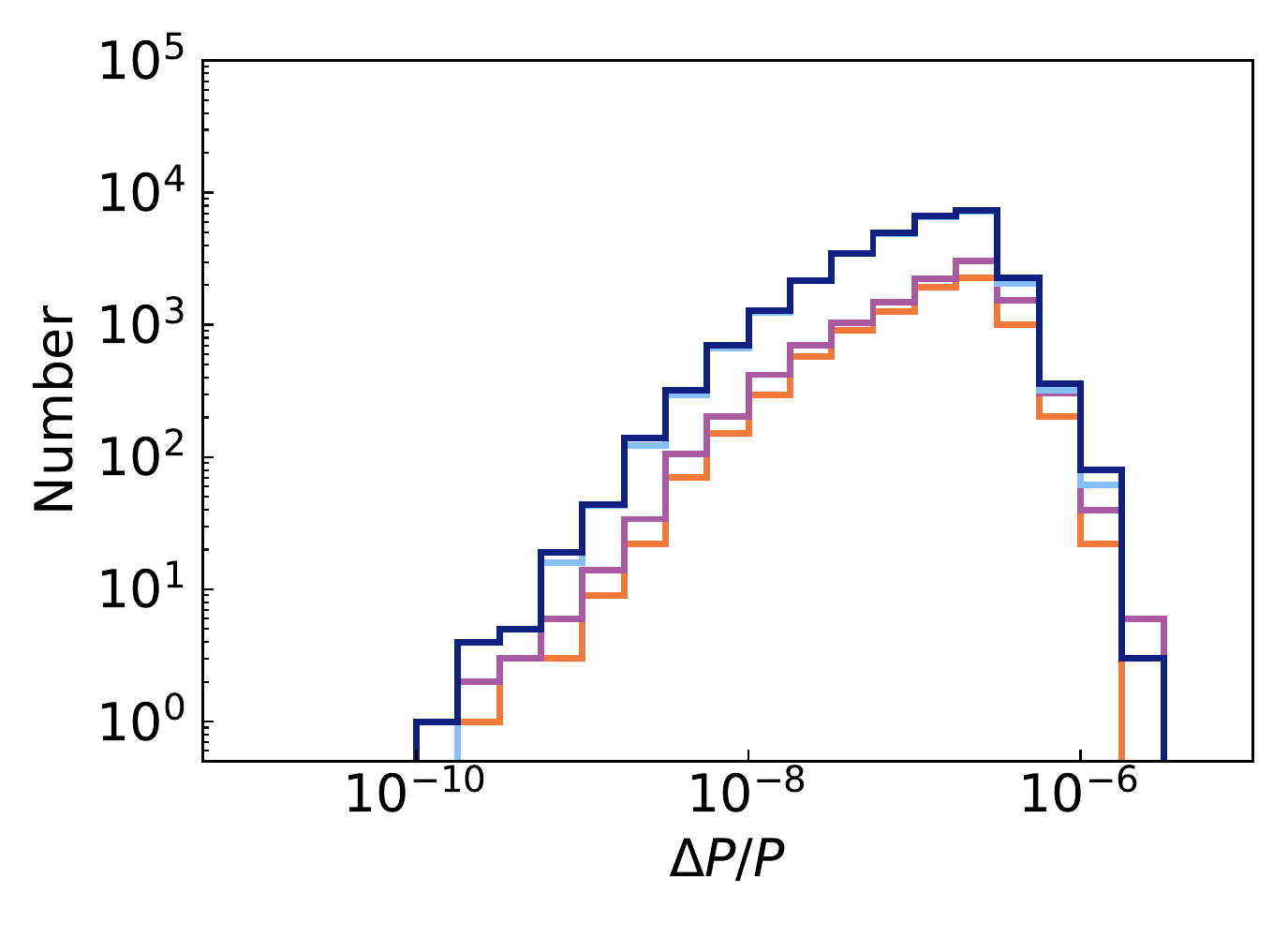}
\end{minipage}

\begin{minipage}{0.5\textwidth}
\includegraphics[width=0.8\textwidth]{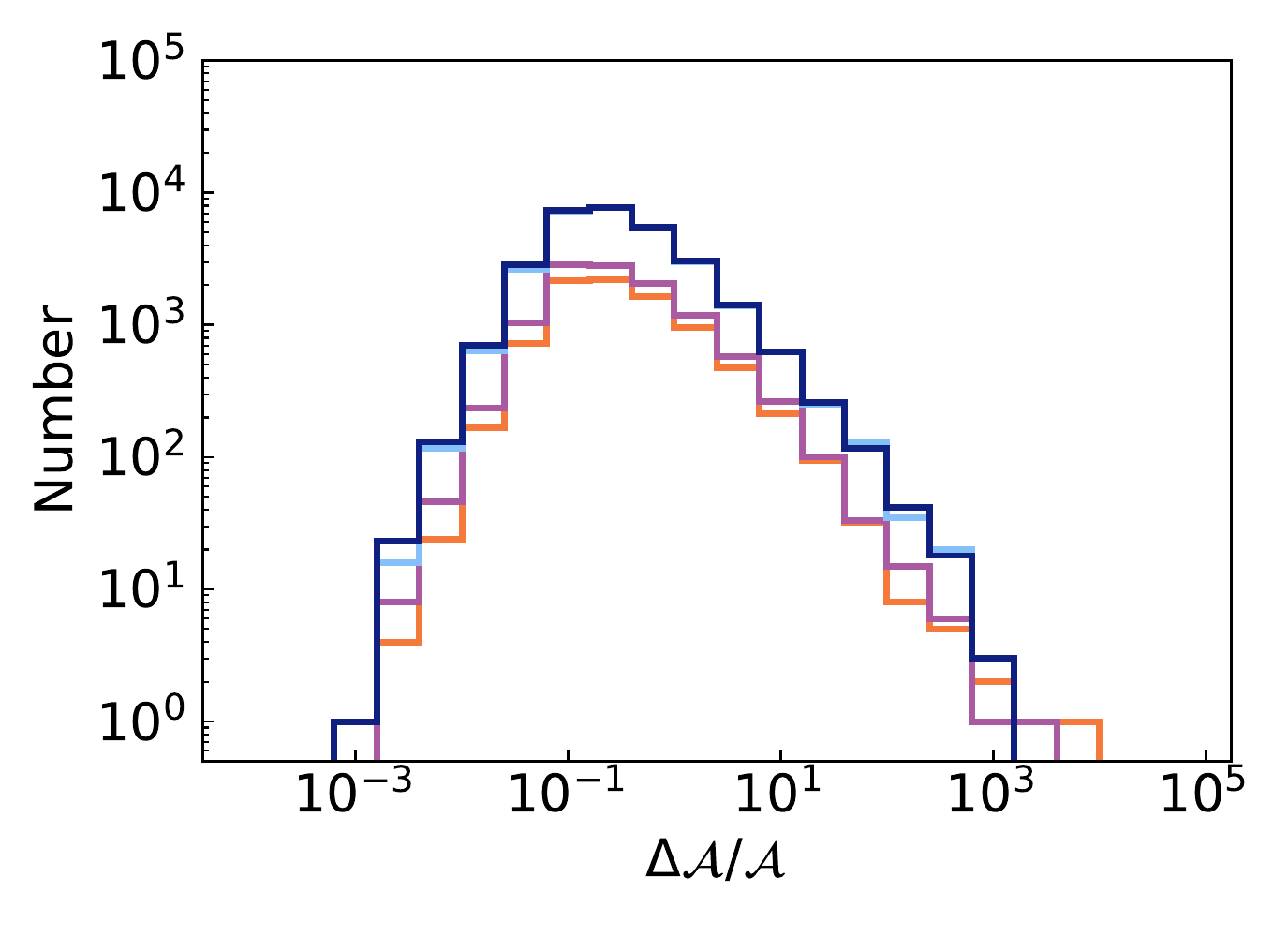}
\end{minipage}%
\begin{minipage}{0.5\textwidth}
\includegraphics[width=0.8\textwidth]{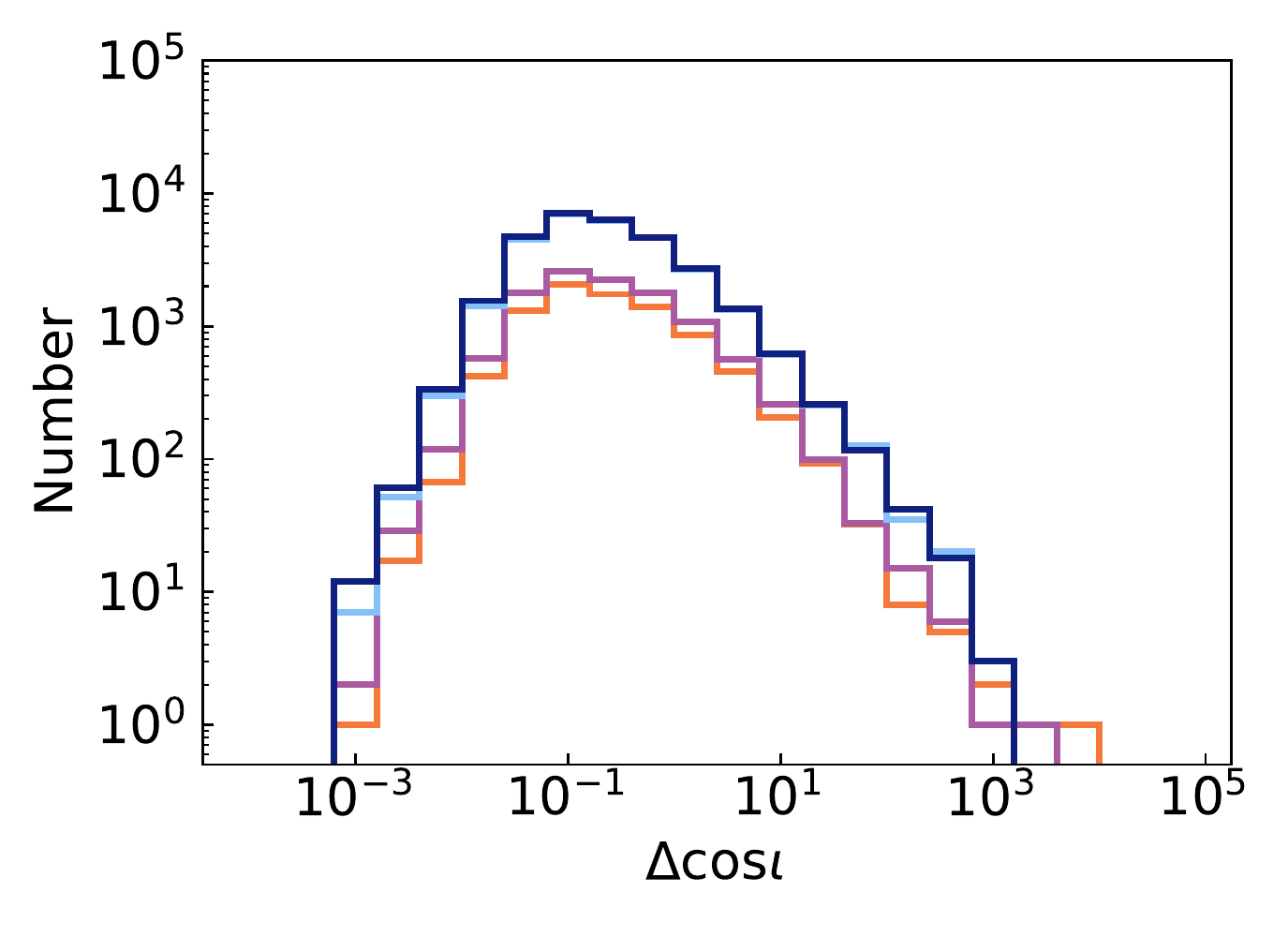}
\end{minipage}

\begin{minipage}{0.5\textwidth}
	\includegraphics[width=0.8\textwidth]{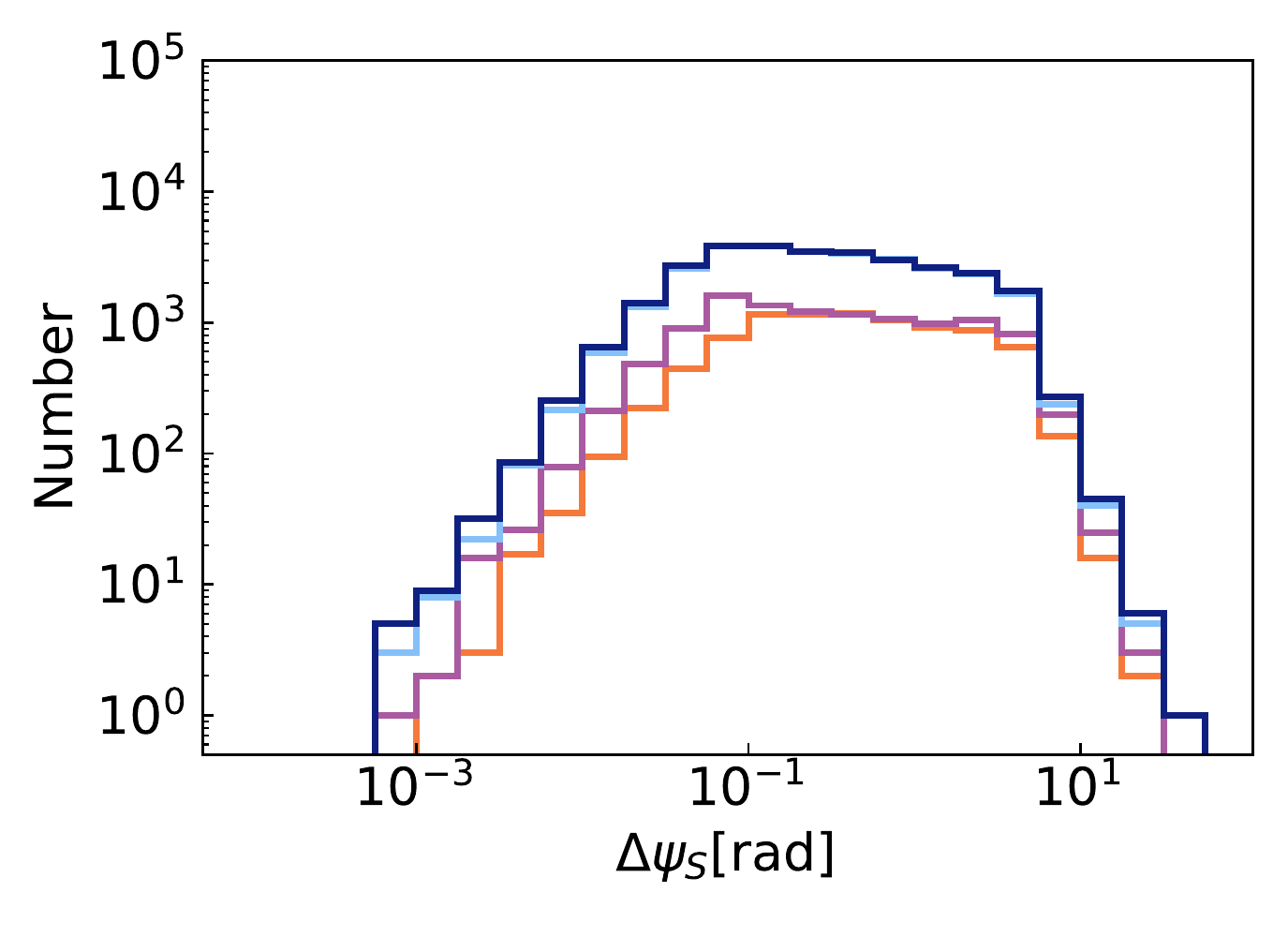}
\end{minipage}%
\begin{minipage}{0.5\textwidth}
	\includegraphics[width=0.8\textwidth]{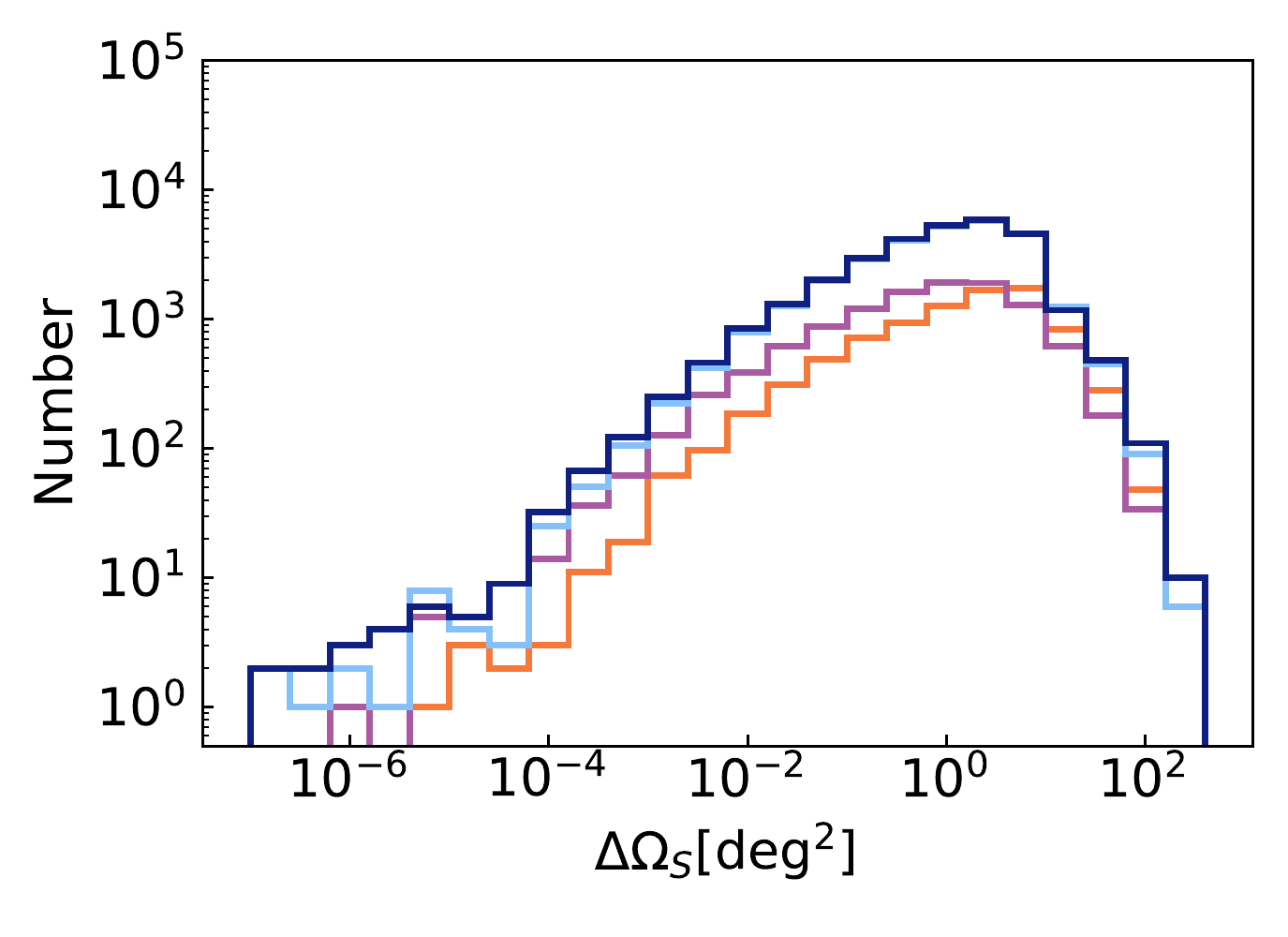}
\end{minipage}
\caption{The histograms (un-normalized) for the \ac{SNR} and the uncertainties of parameter estimation for the resolvable binaries of different detection scenarios with TianQin (orange), TQ I+II (magenta), TQ + LISA (light blue) and TQ I+II + LISA (blue).
}
\label{fig:hist}
\end{figure*}

FIG.~\ref{fig:hist} illustrates the distributions of the SNRs and relative uncertainties on binary parameters $\mathcal{A}$, $P$, $\cos{\iota}$, $\psi_S$, and sky position $\Omega_{\rm S}$ (Eq.~\ref{eq:Omega}).
The figure shows that most sources have a relatively low SNR ($\lesssim$10), and that there is a non-negligible number of sources with SNR $>100$ reaching a maximum of $\sim1000$. 
These high-SNR binaries are also well-localized ones (because $\Delta \Omega_{\rm S} \propto 1/ \rho^2$); theretofore, they will be good candidates for EM follow-up and  multi-messenger studies \cite{Littenberg:2013}.
We find that for $90\%$ of detections, the uncertainty on $\Delta  P/P$ falls within the range $(0.15 - 4.63)\times 10^{-7}$, on $\Delta \mathcal{A}/\mathcal{A}$ within $0.04 - 5.02$, on $\Delta \cos\iota$ within $0.02 - 4.95$, on $\Delta \psi_S$ within $0.03 - 4.01$ rad, and on $\Delta \Omega_{\rm S}$ within $0.02 - 21.36$ deg$^2$.
The median values of these uncertainties are: $\Delta  P/P = 1.41 \times 10^{-7}$, $\Delta \mathcal{A}/\mathcal{A} = 0.26$, $\Delta \cos\iota = 0.20$, $\Delta \psi_S = 0.39$ rad, and $\Delta \Omega_{\rm S} = 1.85$ deg$^2$.
We highlight that TianQin (TQ) can locate 39\% of \acp{DWD} to within better than 1\,deg$^2$, while TQ I+II can locate 54\% of detections within 1\,deg$^2$.

Next, we explore the additional cases of TianQin operating in combination with LISA\footnote{For LISA, we adopt the sensitivity curve from \cite{Robson:2019}.}: TQ + LISA and TQ I+II + LISA.
For these additional cases, the mission lifetimes of TQ and TQ I+II are assumed to be 5 years, while that for LISA is taken to be 4 years \citep{LISA:2017}. 
We verify that by adding LISA to the network, the total number of detected \acp{DWD} doubles.
This is due to the fact that LISA is sensitive to relatively lower \ac{GW} frequencies, where the number of \acp{DWD} is larger. 

\begin{figure*}
	\centering
	\begin{minipage}{0.5\textwidth}
		\includegraphics[width=0.8\textwidth]{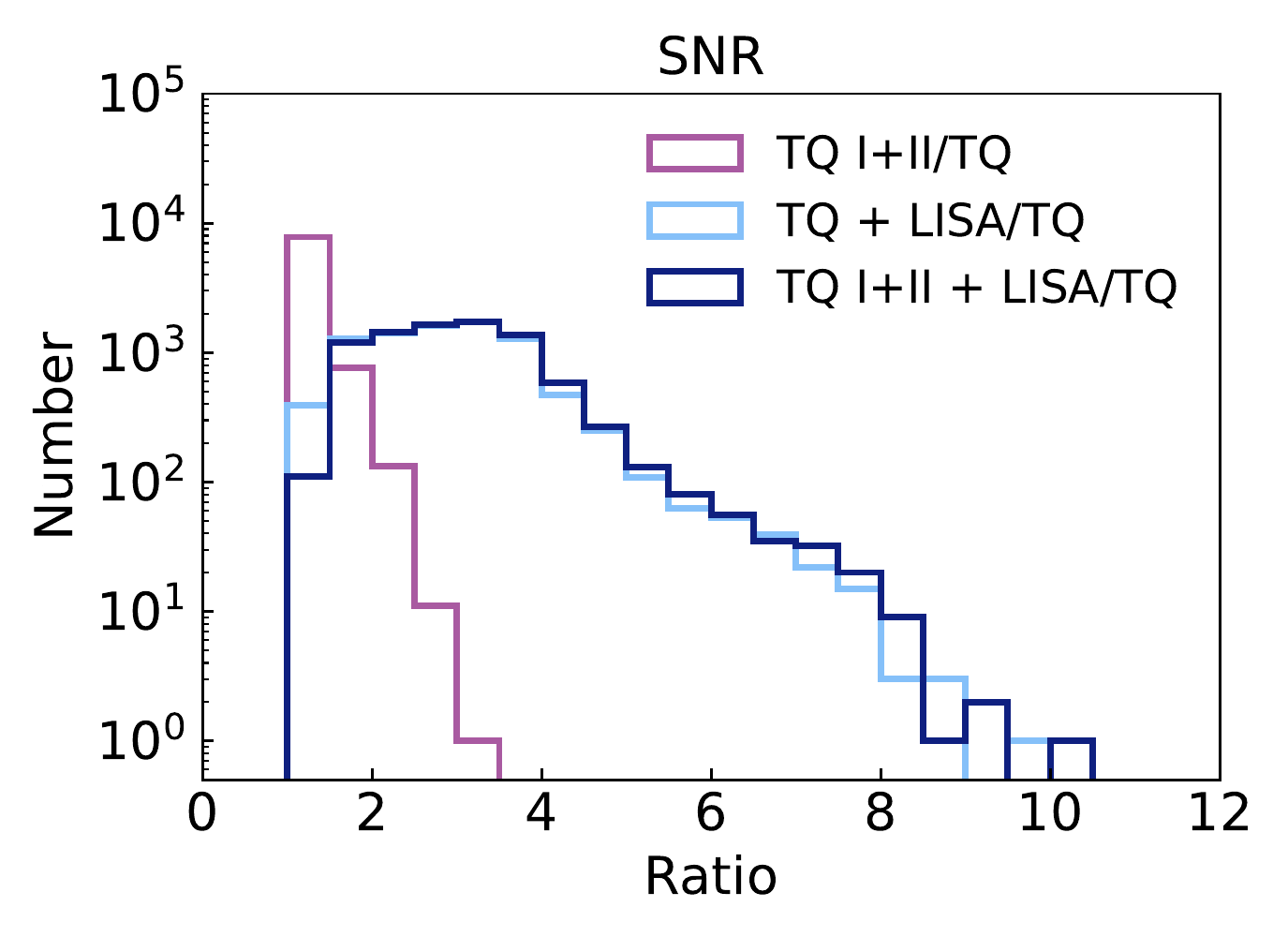}
	\end{minipage}%
	\begin{minipage}{0.5\textwidth}
		\includegraphics[width=0.8\textwidth]{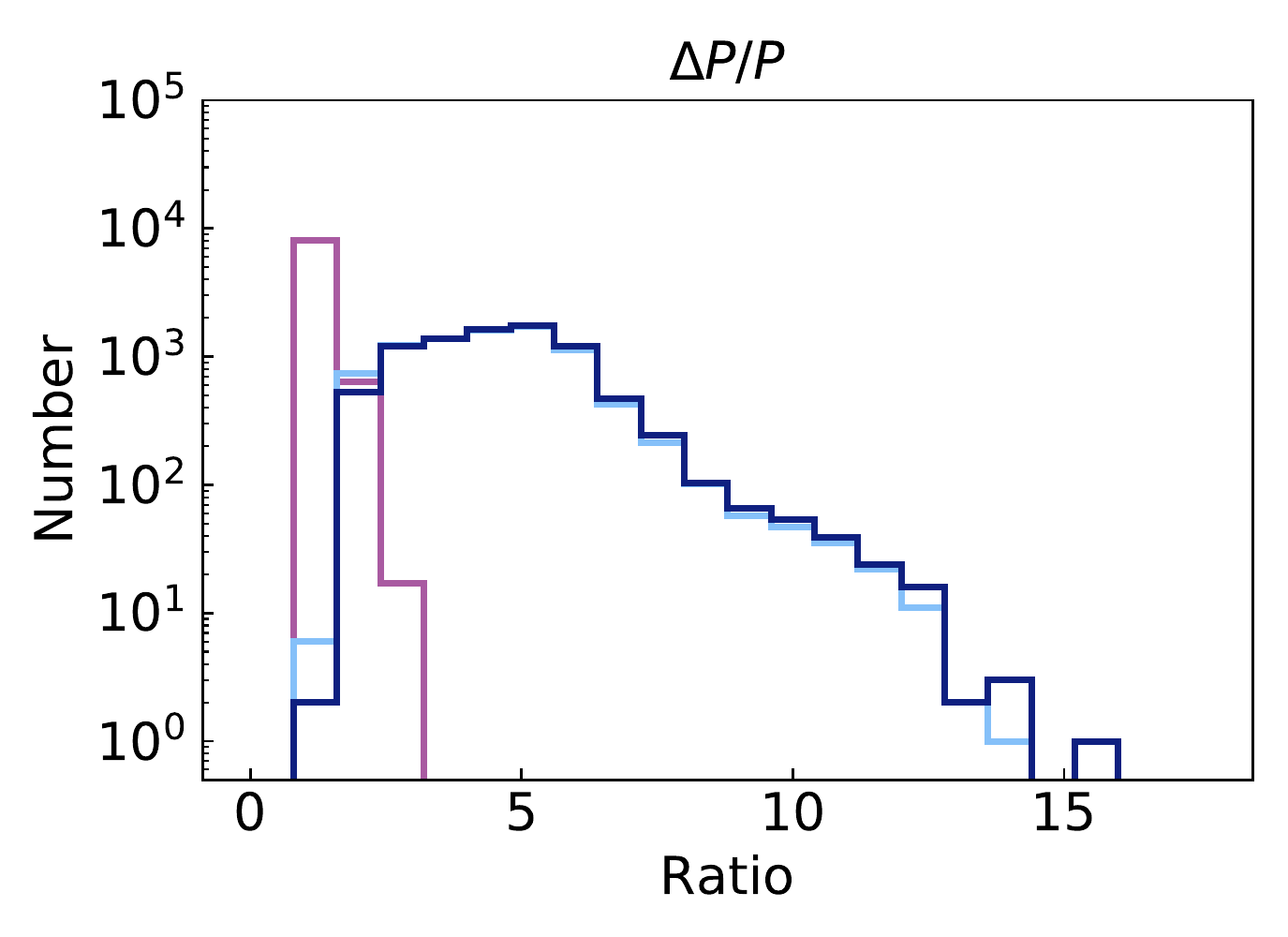}
	\end{minipage}

	\begin{minipage}{0.5\textwidth}
		\includegraphics[width=0.8\textwidth]{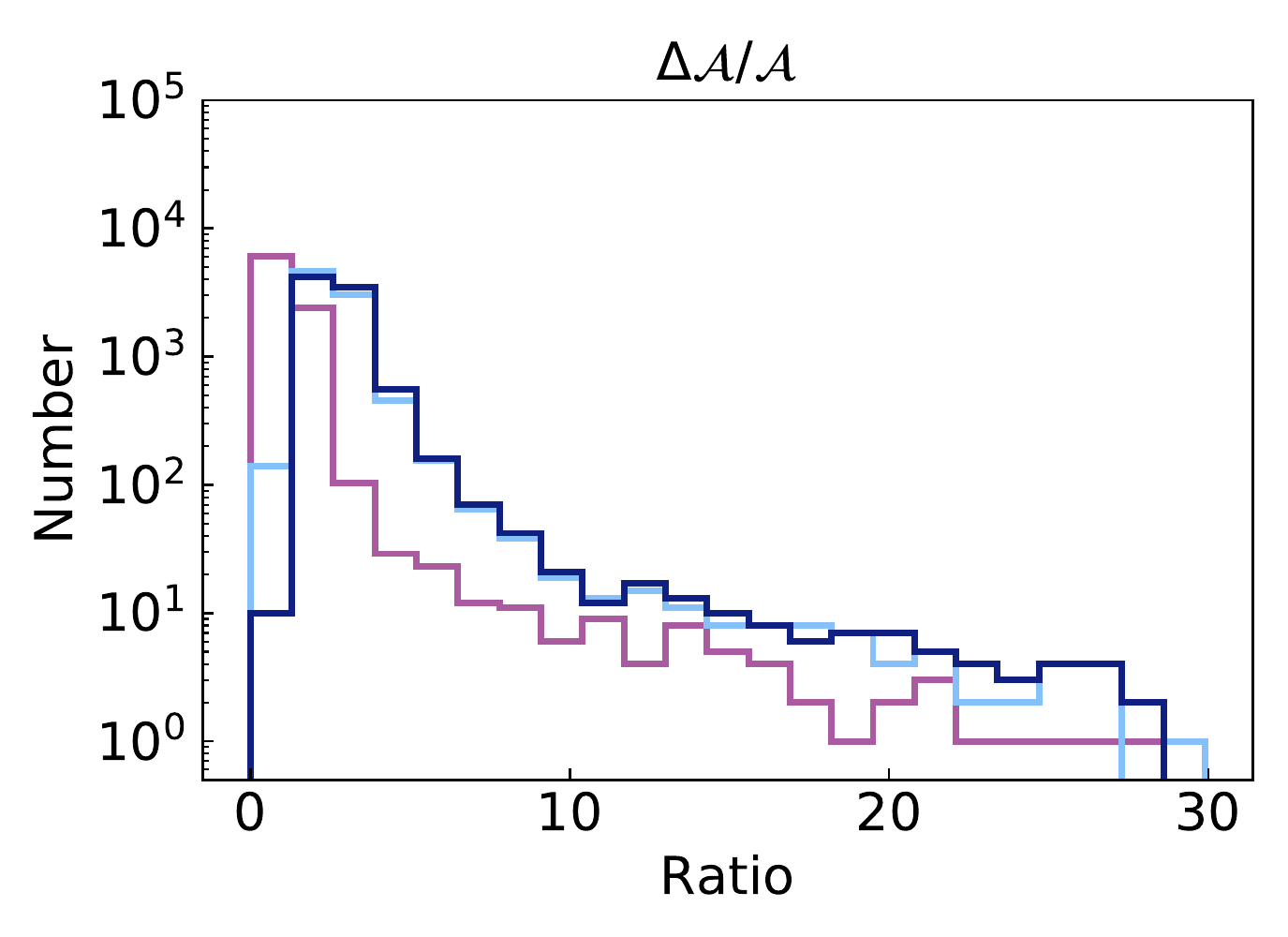}
	\end{minipage}%
	\begin{minipage}{0.5\textwidth}
		\includegraphics[width=0.8\textwidth]{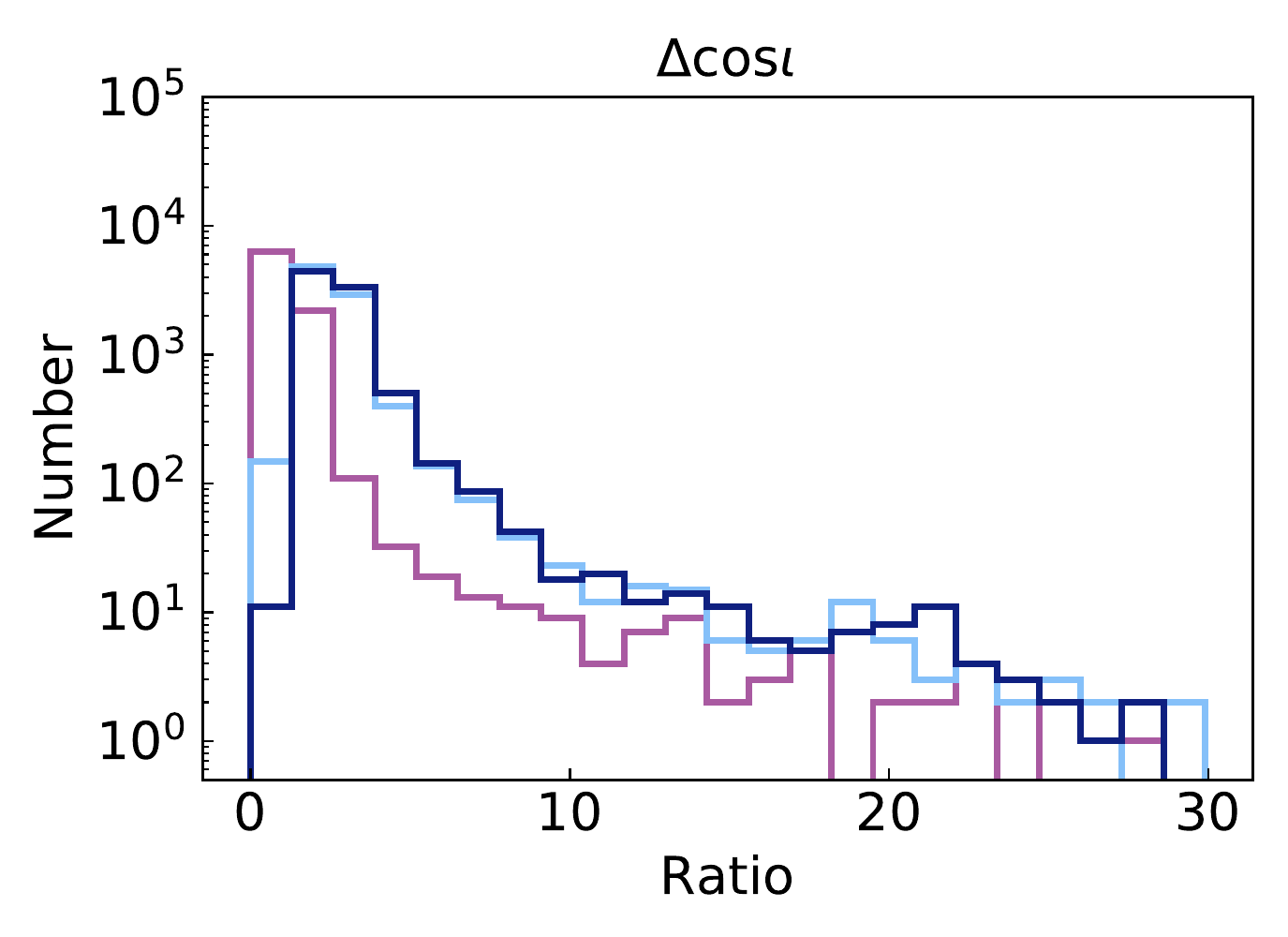}
	\end{minipage}

	\begin{minipage}{0.5\textwidth}
		\includegraphics[width=0.8\textwidth]{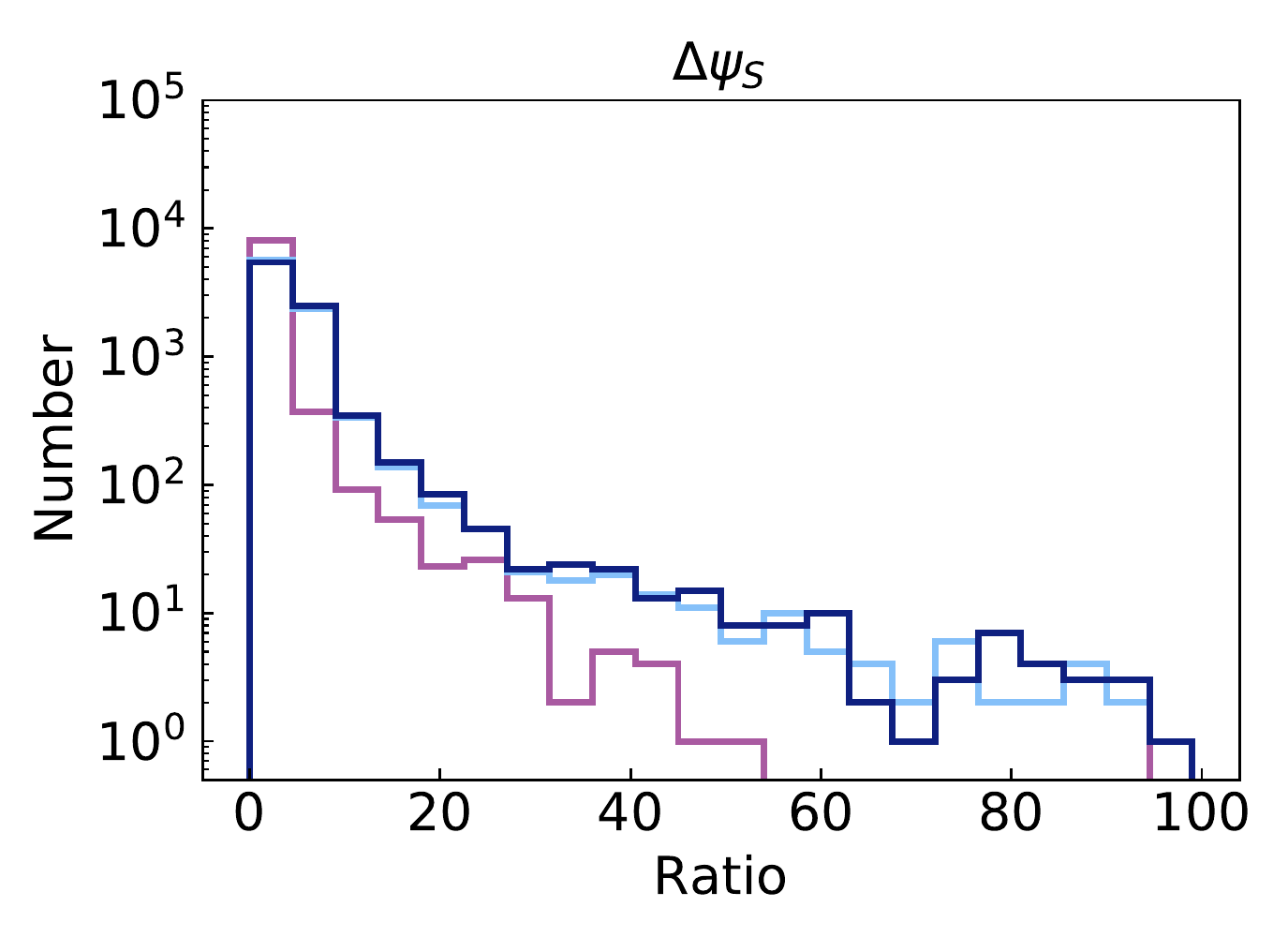}
	\end{minipage}%
	\begin{minipage}{0.5\textwidth}
		\includegraphics[width=0.8\textwidth]{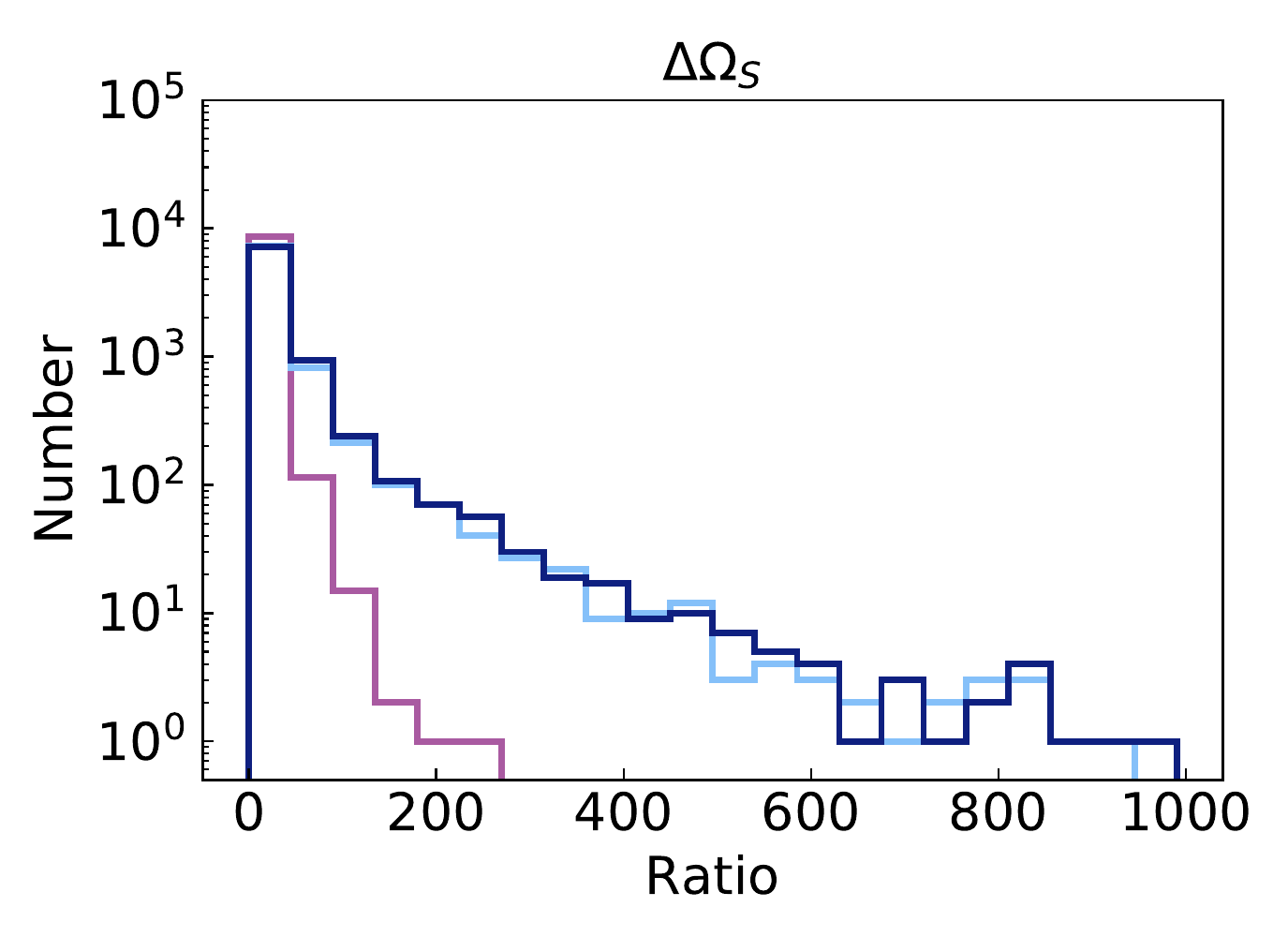}
	\end{minipage}
	\caption{Histograms showing improvement of parameter estimation uncertainties with repect to TianQin (TQ). The horizontal axis shows the ratio of parameter uncertainties between TQ and the corresponding network, the larger value represents better improvement.}
	\label{fig:hist_ratio}
\end{figure*}

As shown in Eq.~(\ref{eq:multi}), an additional detector can increase the SNR of a source, and parameter estimation can also benefit.
We also look at the improvement in the parameter estimation precision for the 8710 resolvable binaries for TianQin.
In FIG.~\ref{fig:hist_ratio}, we present the histograms of the ratio between the uncertainties when measured by TianQin alone, and when measured by a network of detectors.
The top left panel of FIG.~\ref{fig:hist_ratio} shows that the improvement on the SNR is within a factor of 10, while the improvements on the parameters uncertainties are within a factor of a few dozens for $\cos{\iota}$ and $\mathcal{A}$ and  are largely comparable for all three networks.
Improvements on the SNR, $\mathcal{A}$, $P$, $\psi_S$, and $\Omega_{\rm S}$ are larger for TQ + LISA and TQ I+II + LISA; those on $\psi_S$ and $\Omega_{\rm S}$ can reach up to 2 or 3 orders of magnitude.

We remark that (1) TQ + LISA and TQ I+II + LISA are better than TianQin and TQ I+II in determining \acp{DWD}' periods. (2) TianQin and TQ I+II are slightly better than TQ + LISA and TQ I+II + LISA in determining \ac{GW} amplitudes and $\cos(\iota)$. (3) TQ I+II is better than TQ + LISA and TQ I+II + LISA, and the latter two are better than TianQin in determining the sky positions. (4) The result for the polarization angle $\psi_S$ is a bit mixed, but the three networks of detectors usually perform better than TianQin alone.

\subsection{The estimation of the merger rate}
In this section, we estimate the number of \ac{DWD} mergers that can be expected for TianQin.
\acp{DWD} typically merge in the frequency ranging from decihertz to a few hertz. 
Therefore, the inspiral \ac{GW} signals can be detected by TianQin.

We consider a \ac{DWD} with equal mass components of 1M$_\odot$, so that the total mass of the binary is larger the Chandrasekhar mass limit. 
We model its chirping signal with the IMRPhenomPv2 waveform \cite{Hannam:2014} and calculate SNR using Eq.~(\ref{eq:inner product}). 
Following \citet{Wang:2019} and assuming a mission lifetime of 5 years for TianQin, we find that the SNR of our example \ac{DWD} binary is
\begin{equation}
\rho \approx 20\left(\frac{{\rm 1~ Mpc}}{d}\right)\,\label{DWD-SNR}.
\end{equation}
This result implies that TianQin can detect \ac{SNe Ia} explosions within the virial radius of the Local Group.

The \ac{SNe Ia} rate in the Milky Way is 0.01-0.005/yr \cite{Hallakoun:2019}, and the \ac{DWD} merger rate is 4.5-7 times the \ac{SNe Ia} rate (as most \acp{DWD} would not exceed the Chandrasekhar limit) \cite{Maoz:2018,2020arXiv200501880W}.
This means that an optimistic estimation of the \ac{DWD} merger rate is $\sim$0.07/yr in the Galaxy.

To estimate the \ac{DWD} merger rate in the Local Group, we note that the Local Group is consists of about 60 galaxies, most with masses $<10^8$M$_\odot$. 
Therefore, the total mass of the Local Group galaxies is dominated by the Milky Way and the Andromeda Galaxy \cite{Karachentsev:2006}.
The masses of the Milky Way and the Andromeda Galaxy are $0.8-1.5\times10^{12}\,\mathrm{M}_\odot$ and $1\sim2\times10^{12}\, \mathrm{M}_\odot$ \cite{Karachentsev:2006}, respectively.
Assuming that the \ac{DWD} merger rate is proportional to the galaxy mass, one can obtain that the \ac{DWD} merger rate within the Local Group ranges from 0.0375/yr to 0.25/yr, using the relation
\be R_{\mathrm{total}}\Big(1+\frac{M_{31}}{M_{\mathrm{MW}}}\Big)\times R_{\mathrm{MW}}, \ee
where $M_{31}$ and $M_{\mathrm{MW}}$ are  masses of Andromeda Galaxy and Milky Way, and $R_{\mathrm{MW}}$ is the \ac{DWD} merger rate in our Galaxy.
Therefore, in the optimistic case, TianQin would be able to observe one \ac{DWD} merger event with its lifetime of 5 years.

\section{Summary and discussion}\label{sec:5}
In this paper, we carried out the first prediction for the detection of Galactic \acp{DWD} with TianQin.
For this purpose, we adopted a catalogue of known \acp{DWD} discovered with EM observations and a mock Galactic population constructed using a binary population synthesis method. 
We outlined analytical expressions and numerical methods for computing noise curves, \ac{SNR}, and uncertainties on the measured parameters of monochromatic \ac{GW} sources for the TianQin mission with fixed orientation.
By considering different detector orientations, in this work we also addressed an interesting open question regarding the optimal orientation of the mission. 

First, we assessed the strength of the foreground arising from unresolved Galactic \acp{DWD}. We found that its effect can be largely ignored for the present design sensitivity of the TianQin detector.

When considering the sample of known \acp{DWD}, we found that out of 81 \ac{CVBs} with orbital periods $\lesssim5$ hour, TianQin can detect 12 with SNR $\ge5$ within 5 years of mission lifetime. 
In particular, we found that TianQin will be able to detect J0806 (its main verification source) already after two days of observations. 
We estimated that the expected uncertainty on \ac{GW} amplitude for \vb~is within a few per cent.
For \vb~with small inclination angles (nearly face-on), this uncertainty can be improved by up to a factor of 16, if the binary inclination angle is known {\it a priori}.

When analyzing a synthetic Galactic population of DWD, we found that the overall number of detections is expected to be  $8.7\times10^3$ for the full mission duration of 5 years.
We found typical value (median) of $\sim10^{-7}$ on the relative uncertainty of \acp{DWD}' orbital periods, 0.26 on the relative uncertainty of \ac{GW} amplitude, 0.20 uncertainty on $\cos{\iota}$, and $\sim1$ deg$^2$ uncertainty on sky positions.
About 39\% can be localized to within better than 1 deg$^2$.

Finally, we outlined a proof-of-principle calculation showing that TianQin is expected to detect one \ac{DWD} merger event with a supernovae type Ia-like counterpart during its five years of operation time.

In addition to TianQin's nominal orientation (TQ, pointing towards J0806), we also analyzed a variation of the mission oriented perpendicularly (TQ II), and different networks of simultaneously operational \ac{GW} detectors TQ I+II, TQ + LISA, and TQ I+II + LISA.
Although TQ II and TQ I+II can detect the same set of 12 \vb~as TianQin (TQ), the total number of detections increases by $\sim1.3$ when considering the network TQ I+II.
In addition, the total number of binaries localized to better than 1\,deg$^2$ also increases to 54\% of the total detected sample.
We find that the major advantage of combining TianQin and LISA, besides increasing the total number of detections, consists in the improvement on binary parameter uncertainties by $1$-$2$ orders of magnitude, while the improvement the sky localization can reach up to 3 orders of magnitude.

We are living in the era of large astronomical surveys with the number of known \acp{DWD} increasing every year thanks to surveys like ELM \cite{Brown:2016b} and ZTF \cite{Bellm:2014}.
The upcoming LSST \cite{LSST:2009}, GOTO \cite{Steeghs:2017}, and BlackGem \cite{Bloemen:2015} will further enlarge the sample by the time TianQin will fly.
We show that the TianQin mission has the potential to push the \ac{DWD} field in the regime of robust statistical studies by increasing the number of detected \acp{DWD} to several thousand. 
By combining data from \ac{GW} observatories such as TianQin with those from the aforementioned large optical surveys, we will enable multi-messenger studies and advance our knowledge about these unique binary systems.

\begin{acknowledgments}
We would like to thank Gijs Nelemans, Yan Wang, Jian-dong Zhang, Xin-Chun Hu, Xiao-Hong Li, and Shuxu Yi for helpful comments and discussions.
This work was supported in part by the National Natural Science Foundation of China (Grants No.\,11703098, No.\,91636111, No.\,11690022, No.\,1335012, No.\,11325522, No.\,11735001, No.\,11847241, No.\,11947210, No.\,11673031, No.\,11690024) and the Guangdong Major Project of Basic and Applied Basic Research (Contract No.\,2019B030302001).
VK acknowledges support from the Netherlands Research Council NWO (Rubicon Grants No.\,019.183EN.015).
\end{acknowledgments}

\begin{widetext}

\appendix
\section{Table of the selected candidate verification binaries}\label{app:table-78source}

All the selected \ac{CVBs} are listed in Table \ref{tb:VB_parameter}, with the ecliptic coordinates ($\lambda$, $\beta$); the \ac{GW} frequency $f=2/P$, with $P$ being the orbital period of the corresponding binary stars; the luminosity distance $d$; the inclination angle $\iota$ of the source; and the heavier and lighter masses, $M$ and $m$, respectively, of the component stars. 
In some cases, there is no direct measurement on the masses or the inclination angles, so estimated values are assigned based on the evolutionary stage and the mass ratio of the corresponding system.
All such values are given within square brackets.
We make a conservative choice of 5 kpc for the distance to J0806 \cite{Roelofs:2010}. 
The right column of Table \ref{tb:VB_parameter} uses Roman numerals to denote the sources from which the parameters of the listed sources are taken: (i) \cite{Kupfer:2018}, (ii) \cite{Ramsay:2018}, (iii) \cite{Korol:2017}, (iv) \cite{Burdge:2019a}, (v) \cite{Brown:2020}, (vi) \cite{Burdge:2019b}, (vii) \cite{Coughlin:2020}, (viii) \cite{Brown:2016b}, (ix) \cite{Nelemans:2010}, and the references therein. \\

\renewcommand\arraystretch{1}

\begin{longtable}[h]{c cccccccc}
  \caption{The sample of candidate \vb. 
  \\
  \footnotesize{$^a$ As these systems have no measured parallaxes
  		 form Gaia DR2, the distance is estimated by other previously observations.}}\label{tb:VB_parameter}\\
  \hline
		Source      &$\lambda$&  $\beta$&  $f$&   $d$&   $M$&     $m$&  $\iota$& Refs.     \\
		&    [deg]&    [deg]&[mHz]& [kpc]&[$\mathrm{M}_\odot$]& [$\mathrm{M}_\odot$]& [deg]&\\
  \hline
  \hline
  \endfirsthead
  \multicolumn{9}{c}%
  {\tablename\ \thetable\ -- \textit{Continued from previous page}} \\
  \hline
		Source      &$\lambda$&  $\beta$&  $f$&   $d$&   $M$&     $m$&  $\iota$& Refs.     \\
		&    [deg]&    [deg]&[mHz]& [kpc]&[$\mathrm{M}_\odot$]& [$\mathrm{M}_\odot$]& [deg]&\\
  \hline
  \hline
  \endhead
  \hline \multicolumn{9}{r}{\textit{Continued on next page}} \\
  \endfoot
  \hline
  \endlastfoot

                \multicolumn{9}{c}{$\mathbf{AM\:CVn\:type\:systems}$}\\
		J0806                    & 120.4425&  -4.7040& 6.22&   [5]$^a$&  0.55&    0.27&   38& i\\ 
		V407 Vul                 & 294.9945&  46.7829& 3.51& 1.786& [0.8]& [0.177]& [60]& i\\
		ES Cet                   &  24.6120& -20.3339& 3.22& 1.584& [0.8]& [0.161]& [60]& i\\
		SDSS J135154.46--064309.0& 208.3879&   4.4721& 2.12& 1.317& [0.8]& [0.100]& [60]& i\\
		AM CVn                   & 170.3858&  37.4427& 1.94& 0.299&  0.68&   0.125&   43& i\\
		SDSS J190817.07+394036.4 & 298.2172&  61.4542& 1.84& 1.044& [0.8]& [0.085]&   15& i\\
		HP Lib                   & 235.0882&   4.9597& 1.81& 0.276& 0.645&   0.068&   30& i\\
		PTF1 J191905.19+481506.2 & 309.0023&  69.0290& 1.48& 1.338& [0.8]& [0.066]& [60]& i\\
		ASASSN-14cc              & 303.9576& -42.8640& 1.48& 1.019& [0.6]&  [0.01]& [60]& ii\\
		CXOGBS J175107.6--294037 & 268.0614&  -6.2526& 1.45& 0.971& [0.8]& [0.064]& [60]& i\\
		CR Boo                   & 202.2728&  17.8971& 1.36& 0.337$^a$& 0.885&   0.066&   30& i\\
		KL Dra                   & 334.1334&  78.3217& 1.33& 0.956&  0.76&   0.057& [60]& ii,iii\\
		V803 Cen                 & 216.1673& -30.3166& 1.25& 0.347$^a$& 0.975&   0.084& 13.5& i\\
		PTF1 J071912.13+485834.0 & 104.3883&  26.5213& 1.24& 0.861& [0.8]& [0.053]& [60]& i,ii\\
		SDSS J092638.71+362402.4 & 132.2867&  20.2342& 1.18& 0.577&  0.85&   0.035& 82.6& ii,iii\\
		CP Eri                   &  42.1327& -26.4276& 1.17& 0.964& [0.8]& [0.049]& [60]& i,ii\\
		SDSS J104325.08+563258.1 & 136.2923&  43.9158& 1.17& 0.979& [0.6]&  [0.01]& [60]& ii\\
		CRTS J0910-2008          & 147.3411& -34.5979& 1.12& 1.113& [0.6]&  [0.01]& [60]& ii\\
		CRTS J0105+1903          &  22.5049&  11.1283& 1.05& 0.734& [0.6]&  [0.01]& [60]& ii\\
		V406 Hya/2003aw          & 140.7336& -21.2342& 0.99& 0.504& [0.8]& [0.040]& [60]& i,ii\\
		SDSS J173047.59+554518.5 & 248.6846&  78.6529& 0.95& 0.911& [0.6]&  [0.01]& [60]& ii\\
		2QZ J142701.6--012310    & 214.8878&  12.4608& 0.91& 0.677& [0.6]& [0.015]& [60]& ii,iii\\
		SDSS J124058.03--015919.2& 190.1933&   2.2262& 0.89& 0.577& [0.8]& [0.035]& [60]& i,ii\\
		NSV1440                  & 283.1788& -72.6108& 0.89& 0.377& [0.6]&  [0.01]& [60]& ii\\
		SDSS J012940.05+384210.4 &  35.8760&  27.0847& 0.89& 0.508& [0.8]& [0.034]& [60]& i,ii\\
		SDSS J172102.48+273301.2 & 256.3525&  50.5292& 0.87& 0.995& [0.6]&  [0.01]& [60]& ii\\
		ASASSN-14mv              & 107.1184&  -1.4233& 0.82& 0.247& [0.6]&  [0.01]& [60]& ii\\
		ASASSN-14ei              &  15.2639& -59.9095& 0.78& 0.255& [0.6]&  [0.01]& [60]& ii\\
		SDSS J152509.57+360054.5 & 214.3101&  52.2364& 0.75& 0.524& [0.6]&  [0.01]& [60]& ii\\
		SDSS J080449.49+161624.8 & 119.9052&  -3.9847& 0.75& 0.828& [0.8]& [0.027]& [60]& i,ii\\
		SDSS J141118.31+481257.6 & 183.5559&  55.8748& 0.72& 0.429& [0.6]&  [0.01]& [60]& ii\\
		GP Com                   & 187.7210&  23.0012& 0.72& 0.073&  0.59&   0.011& [60]& ii,iii\\
		SDSS J090221.35+381941.9 & 126.7527&  20.5254& 0.69& 0.461& [0.6]&  [0.01]& [60]& ii\\
		ASASSN-14cn              & 183.9170&  78.0916& 0.67& 0.259&  0.87&   0.025& 86.3& i,ii\\
		SDSS J120841.96+355025.2 & 165.8193&  33.3289& 0.63& 0.202& [0.8]& [0.022]& [60]& i,ii\\
		SDSS J164228.06+193410.0 & 245.3756&  41.3659& 0.62& 1.044& [0.6]&  [0.01]& [60]& ii\\
		SDSS J155252.48+320150.9 & 225.2376&  50.6483& 0.59& 0.443& [0.6]&  [0.01]& [60]& ii\\
		SDSS J113732.32+405458.3 & 156.4126&  34.8546& 0.56& 0.209& [0.6]&  [0.01]& [60]& ii\\
		V396 Hya/CE 315          & 205.7504& -14.4638& 0.51& 0.094& [0.8]& [0.016]& [60]& i,ii\\
		SDSS J1319+5915          & 159.2931&  59.0926& 0.51& 0.205& [0.6]&  [0.01]& [60]& ii\\
                \multicolumn{9}{c}{$\mathbf{detached\:DWD}$}\\
        ZTF J153932.16+502738.8  & 205.0315&  66.1616& 4.82& 1.262&  0.61&    0.21&   84& iv\\
		SDSS J065133.34+284423.4 & 101.3396&   5.8048& 2.61& 0.933& 0.247&    0.49& 86.9& i\\
		SDSS J093506.92+441107.0 & 130.9795&  28.0912& 1.68& 0.645$^a$& 0.312&    0.75& [60]& i\\
		SDSS J232230.20+050942.06& 353.4373&   8.4572& 1.66& 0.779&  0.24&    0.27&   27& v\\
		PTF J053332.05+020911.6  &  82.9097& -21.1234& 1.62& 1.253&  0.65&   0.167& 72.8& vi\\
		SDSS J010657.39--100003.3&  11.4582& -15.7928& 0.85& 0.758& 0.188&    0.57&   67& i,iii\\
		SDSS J163030.58+423305.7 & 231.7612&  63.0501& 0.84& 1.019& 0.298&    0.76& [60]& i\\
		SDSS J082239.54+304857.2 & 120.6816&  11.0965& 0.83& 0.861& 0.304&   0.524& 88.1& i,iii\\
		ZTF J190125.42+530929.5  & 306.8131&  74.6335& 0.82& 0.898&  0.50&    0.20& 86.2& vii\\
		SDSS J104336.27+055149.9 & 160.1545&  -2.0480& 0.73& 1.744& 0.183&    0.76& [60]& i,iii\\
		SDSS J105353.89+520031.0 & 141.2200&  40.8002& 0.54& 0.683& 0.204&    0.75& [60]& i,iii\\
		SDSS J005648.23--061141.5&  10.6273& -11.3044& 0.53& 0.620& 0.180&    0.82& [60]& i,iii\\
		SDSS J105611.02+653631.5 & 130.4076&  52.2268& 0.53& 1.104& 0.334&    0.76& [60]& i,iii\\
		SDSS J092345.59+302805.0 & 133.7151&  14.4268& 0.51& 0.299& 0.275&    0.76& [60]& i\\
		SDSS J143633.28+501026.9 & 187.5011&  59.9313& 0.50& 1.011& 0.234&    0.78& [60]& i,iii\\
		SDSS J082511.90+115236.4 & 125.7257&  -7.1746& 0.40& 1.786& 0.278&    0.80& [60]& i,iii\\
		WD 0957--666             & 208.5263& -67.3013& 0.38& 0.163&  0.37&    0.32&   68& i,iii\\
		SDSS J174140.49+652638.7 & 208.8283&  87.8286& 0.38& 1.159& 0.170&    1.17& [60]& i,iii\\ 
		SDSS J075552.40+490627.9 & 110.9953&  27.7583& 0.37& 2.620$^a$& 0.176&    0.81& [60]& iii\\
		SDSS J233821.51--205222.8& 346.5446& -16.9689& 0.30& 0.429&  0.15&   0.263& [60]& iii\\
		SDSS J230919.90+260346.7 & 359.6100&  28.7808& 0.30& 1.765& 0.176&    0.96& [60]& viii\\
		SDSS J084910.13+044528.7 & 133.3917& -12.5404& 0.29& 1.002& 0.176&    0.65& [60]& iii\\
		SDSS J002207.65--101423.5&   0.9548& -11.5858& 0.29& 1.151$^a$&  0.21&   0.375& [60]& iii\\
		SDSS J075141.18--014120.9& 120.3746& -22.2324& 0.29& 1.741&  0.97&   0.194& [60]& iii\\
		SDSS J211921.96--001825.8& 322.1533&  14.5725& 0.27& 1.053&  0.74&   0.158& [60]& iii\\
		SDSS J123410.36--022802.8& 188.8196&   1.1194& 0.25& 0.754&  0.09&    0.23& [60]& iii\\
		SDSS J100559.10+224932.2 & 145.4432&  10.4254& 0.24& 0.555&  0.36&    0.31& 88.9& iii\\
		SDSS J115219.99+024814.4 & 177.1265&   1.8106& 0.23& 0.718&  0.47&    0.41& 89.2& iii\\
		SDSS J105435.78--212155.9& 173.8923& -26.0107& 0.22& 1.313&  0.39&   0.168& [60]& iii\\
		SDSS J074511.56+194926.5 & 114.6397&  -1.3939& 0.20& 0.875&   0.1&   0.156& [60]& iii\\
		WD 1242--105             & 194.5586&  -5.5520& 0.19& 0.040&  0.56&    0.39& 45.1& i,iii\\
		SDSS J110815.50+151246.6 & 162.1662&   8.9070& 0.19& 0.698$^a$&  0.42&   0.167& [60]& iii\\
		WD 1101+364	             & 152.2513&  27.6895& 0.16& 0.088&  0.36&    0.31& [60]& iii\\
		WD 1704+4807BC           & 242.3234&  70.1865& 0.16& 0.039&  0.39&    0.56& [60]& ix\\
		SDSS J011210.25+183503.7 &  23.7268&  10.1149& 0.16& 0.843&  0.62&    0.16& [60]& iii\\
		SDSS J123316.20+160204.6 & 181.0654&  17.9826& 0.15& 1.207& 0.169&    0.98& [60]& viii\\
		SDSS J113017.42+385549.9 & 156.0760&  32.4474& 0.15& 0.884&  0.72&   0.286& [60]& iii\\
		SDSS J111215.82+111745.0 & 164.6171&   5.6844& 0.13& 0.384&  0.14&   0.169& [60]& iii\\
		SDSS J100554.05+355014.2 & 140.4558&  22.5307& 0.13& 1.747& 0.168&    0.75& [60]& viii\\
		SDSS J144342.74+150938.6 & 213.1397&  29.4368& 0.12& 0.839&  0.84&   0.181& [60]& iii\\
		SDSS J184037.78+642312.3 & 337.4095&  85.2636& 0.12& 0.829&  0.65&   0.177& [60]& iii\\

\end{longtable}

\vspace{1cm}

\section{SNR OF CANDIDATE  VERIFICATION BINARIES}
The GW amplitudes and SNR of all selected CVBs are listed in Table \ref{tb:VB_SNR}, assuming a nominal mission lifetime of five years and the three configurations of TianQin, $\phi_0=\pi$ and $\psi_S=\pi/2$ for all binaries.

\renewcommand\arraystretch{1}

\begin{longtable}[h]{cccccc}
  \caption{The expected amplitude $\mathcal{A}$ and \ac{SNR} of 81 candidate verification binaries. $\mathcal{A}$ is given in units of $10^{-23}$. 
  	} 
  \label{tb:VB_SNR}\\
  \hline
  Source     & $\mathcal{A}$&                &        SNR      &                    \\
           &              & TQ& TQ II& TQ I+II \\
  \hline
  \hline
  \endfirsthead
  \multicolumn{5}{c}%
  {\tablename\ \thetable\ -- \textit{Continued from previous page}} \\
  \hline
  Source     & $\mathcal{A}$&                &        SNR      &                    \\
           &              & TQ& TQ II& TQ I+II \\
  \hline
  \hline
  \endhead
  \hline \multicolumn{5}{r}{\textit{Continued on next page}} \\
  \endfoot
  \hline
  \endlastfoot

  \multicolumn{5}{c}{$\mathbf{AM\:CVn\:type\:systems}$}\\
 J0806                    &  6.4& 116.202&  41.657& 123.443\\      
 V407 Vul                 & 11.0&  41.528&  21.537&  46.780\\ 
 ES Cet                   & 10.7&  17.775&  42.110&  45.708\\ 
 SDSS J135154.46--064309.0&  6.2&   4.454&  11.345&  12.188\\ 
 AM CVn                   & 28.3&  31.245&  37.499&  48.810\\ 
 SDSS J190817.07+394036.4 &  6.1&   8.622&   5.077&  10.006\\ 
 HP Lib                   & 15.7&  16.619&  29.427&  33.795\\ 
 PTF1 J191905.19+481506.2 &  3.2&   1.526&   1.122&   1.894\\ 
 ASASSN-14cc              &  0.5&   0.338&   0.188&   0.387\\ 
 CXOGBS J175107.6--294037 &  4.2&   3.022&   2.172&   3.722\\ 
 CR Boo                   & 12.9&   5.473&  14.029&  15.058\\ 
 KL Dra                   &  3.5&   1.109&   1.006&   1.497\\ 
 V803 Cen                 & 16.0&   6.187&  15.026&  16.249\\ 
 PTF1 J071912.13+485834.0 &  3.6&   1.844&   0.982&   2.089\\ 
 SDSS J092638.71+362402.4 &  3.6&   1.175&   0.664&   1.350\\ 
 CP Eri                   &  2.8&   0.676&   1.384&   1.540\\ 
 SDSS J104325.08+563258.1 &  0.5&   0.178&   0.112&   0.211\\ 
 CRTS J0910-2008          &  0.4&   0.164&   0.104&   0.194\\ 
 CRTS J0105+1903          &  0.6&   0.108&   0.256&   0.277\\ 
 V406 Hya/2003aw          &  4.0&   1.414&   0.745&   1.599\\ 
 SDSS J173047.59+554518.5 &  0.4&   0.069&   0.066&   0.096\\ 
 2QZ J142701.6--012310    &  0.9&   0.115&   0.282&   0.304\\ 
 SDSS J124058.03--015919.2&  2.8&   0.436&   0.842&   0.948\\ 
 NSV1440                  &  1.0&   0.140&   0.130&   0.191\\ 
 SDSS J012940.05+384210.4 &  3.1&   0.389&   0.882&   0.964\\ 
 SDSS J172102.48+273301.2 &  0.4&   0.069&   0.063&   0.093\\ 
 ASASSN-14mv              &  1.5&   0.388&   0.174&   0.425\\ 
 ASASSN-14ei              &  1.4&   0.133&   0.192&   0.233\\ 
 SDSS J152509.57+360054.5 &  0.7&   0.059&   0.097&   0.114\\ 
 SDSS J080449.49+161624.8 &  1.4&   0.304&   0.119&   0.326\\ 
 SDSS J141118.31+481257.6 &  0.8&   0.068&   0.093&   0.115\\ 
 GP Com                   &  5.0&   0.490&   0.882&   1.009\\ 
 SDSS J090221.35+381941.9 &  0.7&   0.121&   0.053&   0.132\\ 
 ASASSN-14cn              &  4.0&   0.220&   0.227&   0.316\\ 
 SDSS J120841.96+355025.2 &  4.1&   0.383&   0.408&   0.560\\ 
 SDSS J164228.06+193410.0 &  0.3&   0.025&   0.029&   0.038\\ 
 SDSS J155252.48+320150.9 &  0.7&   0.039&   0.060&   0.072\\ 
 SDSS J113732.32+405458.3 &  1.4&   0.110&   0.095&   0.145\\ 
 V396 Hya/CE 315          &  5.5&   0.221&   0.535&   0.579\\  
 SDSS J1319+5915          &  1.3&   0.063&   0.062&   0.089\\ 
 \multicolumn{5}{c}{$\mathbf{detached\:DWD}$}\\
 ZTF J153932.16+502738.8  & 18.4&  51.351&  60.184&  79.114\\ 
 SDSS J065133.34+284423.4 & 16.2&  26.535&  15.700&  30.831\\ 
 SDSS J093506.92+441107.0 & 29.9&  28.797&  14.245&  32.128\\
 SDSS J232230.20+050942.06&  8.7&   9.973&  12.371&  15.891\\
 PTF J053332.05+020911.6  &  7.6&   4.965&   4.042&   6.402\\   
 SDSS J010657.39--100003.3&  8.3&   0.989&   1.892&   2.135\\ 
 SDSS J163030.58+423305.7 & 11.6&   1.423&   1.721&   2.233\\ 
 SDSS J082239.54+304857.2 & 10.4&   1.713&   0.887&   1.929\\
 ZTF J190125.42+530929.5  &  6.5&   0.614&   0.549&   0.824\\  
 SDSS J104336.27+055149.9 &  3.9&   0.649&   0.561&   0.857\\ 
 SDSS J105353.89+520031.0 &  9.0&   0.698&   0.457&   0.835\\ 
 SDSS J005648.23--061141.5&  9.3&   0.475&   0.931&   1.045\\ 
 SDSS J105611.02+653631.5 &  8.7&   0.570&   0.378&   0.684\\ 
 SDSS J092345.59+302805.0 & 26.2&   2.422&   1.138&   2.675\\ 
 SDSS J143633.28+501026.9 &  6.7&   0.262&   0.359&   0.444\\ 
 SDSS J082511.90+115236.4 &  3.9&   0.235&   0.094&   0.253\\ 
 WD 0957--666             & 25.7&   0.502&   0.621&   0.798\\ 
 SDSS J174140.49+652638.7 &  4.9&   0.104&   0.103&   0.147\\ 
 SDSS J075552.40+490627.9 &  1.7&   0.072&   0.035&   0.080\\ 
 SDSS J233821.51--205222.8&  3.3&   0.073&   0.078&   0.107\\ 
 SDSS J230919.90+260346.7 &  2.4&   0.046&   0.062&   0.077\\ 
 SDSS J084910.13+044528.7 &  3.2&   0.094&   0.042&   0.103\\ 
 SDSS J002207.65--101423.5&  2.1&   0.036&   0.056&   0.067\\ 
 SDSS J075141.18--014120.9&  2.7&   0.078&   0.032&   0.084\\ 
 SDSS J211921.96--001825.8&  2.9&   0.069&   0.037&   0.078\\ 
 SDSS J123410.36--022802.8&  0.9&   0.011&   0.020&   0.022\\ 
 SDSS J100559.10+224932.2 &  5.3&   0.059&   0.039&   0.071\\ 
 SDSS J115219.99+024814.4 &  6.3&   0.047&   0.062&   0.078\\ 
 SDSS J105435.78--212155.9&  1.3&   0.014&   0.017&   0.022\\ 
 SDSS J074511.56+194926.5 &  0.6&   0.008&   0.003&   0.008\\ 
 WD 1242--105             &109.2&   0.755&   1.656&   1.820\\ 
 SDSS J110815.50+151246.6 &  2.4&   0.022&   0.020&   0.030\\ 
 WD 1101+364	          & 25.4&   0.159&   0.120&   0.199\\ 
 WD 1704+4807BC           & 99.9&   0.376&   0.388&   0.541\\ 
 SDSS J011210.25+183503.7 &  2.2&   0.008&   0.019&   0.020\\ 
 SDSS J123316.20+160204.6 &  2.2&   0.009&   0.014&   0.016\\ 
 SDSS J113017.42+385549.9 &  3.9&   0.019&   0.016&   0.026\\ 
 SDSS J111215.82+111745.0 &  1.4&   0.005&   0.005&   0.008\\ 
 SDSS J100554.05+355014.2 &  1.1&   0.005&   0.003&   0.006\\ 
 SDSS J144342.74+150938.6 &  2.6&   0.005&   0.010&   0.011\\ 
 SDSS J184037.78+642312.3 &  2.1&   0.004&   0.004&   0.005\\ 

\end{longtable}

\vspace{1cm}

\section{Re-expression of the responsed Gravitational Wave Signal}
For convenience of calculation, we rearrange the expression for the waveform in the detector:
	\be h(t) =A(t) \cos\Psi(t)\, ,
	\label{eq:waveform} \ee
	where the waveform amplitude $A(t)$ is
	\be A(t)=\left[(A_+F^+(t))^2 + (A_{\times}F^{\times}(t))^2\right]^{1/2}.
	\label{eq:waveform_amplitude} \ee
	$A_+$ and $A_{\times}$ are given by
	\be A_+ = \mathcal{A}(1+\cos\iota^2),\quad A_{\times} = 2\mathcal{A}\cos\iota.
	\label{} \ee
	The phase of the waveform is
	\be \Psi(t) = 2\pi ft + \phi_0 + \Phi_D(t) + \Phi_P(t),
	\label{eq:waveform_phase} \ee
	The polarization phase $\Phi_P(t)$ is given by
	\be \Phi_P(t) = \tan^{-1}\left(\frac{- A_{\times}F^{\times}(t)}{A_+F^+(t)}\right)\, .
	\label{} \ee

\section{Derivation of the average amplitude}\label{Derivation of average amplitude}
In order to verify our \ac{SNR} calculation, more specifically the calculation of average amplitude, we can obtain the average amplitude from the antenna beam patterns function given by Eq.~(13) in \cite{Hu:2018}:
\begin{eqnarray}
F^+(t,\theta,\phi,\psi) &=& \cos2\psi\xi^+(t;\theta,\phi) -sin2\psi\xi^\times(t;\theta,\phi) \, , \nn\\
F^\times(t,\theta,\phi,\psi) &=& \sin2\psi\xi^+(t;\theta,\phi)+\cos2\psi\xi^\times(t;\theta,\phi) \, ,
\end{eqnarray}
and
\begin{eqnarray}
\xi^+(t;\theta,\phi) &=& \frac{\sqrt{3}}{32} (4\cos2(\kappa-\beta')((3+\cos2\theta)\sin\theta_s\sin 2(\phi-\phi_s)+ 2 \sin(\phi -\phi_s)\sin2\theta\cos\theta _s)  \nn\\
& &-\sin2(\kappa-\beta') (3+\cos2(\phi-\phi_s)(9+\cos2\theta(3-\cos2\theta_s))+6\cos2\theta_s\sin^2(\phi-\phi _s)\nn\\
& &-6\cos2\theta\cos^2\theta_s + 4 \cos(\phi-\phi_s)\sin2\theta\sin2\theta_s )) \, ,\nn\\
\xi^\times(t;\theta,\phi) &=& \frac{\sqrt{3}}{8}(-4\cos2(\kappa-\beta')(\cos2(\phi-\phi_s)\cos\theta\sin\theta_s+\cos(\phi-\phi_s)\sin\theta\cos\theta_s )   \nn\\
& &+\sin2(\kappa-\beta')(\cos\theta(3-\cos2\theta_s)\sin2(\phi_s-\phi)+2\sin(\phi_s-\phi)\sin\theta\sin2\theta_s)) \, .
\end{eqnarray}
where $\kappa=2\pi f_{sc} t+\lambda'$, $f_{sc}\approx 1/(3.65d)$ is the modulation frequency from the rotation of the satellites around the guiding center.
$\lambda'$ and $\beta'$ are some of initial phase of constant.

In the above expression, $\theta=\pi/2-\beta$ and $\phi=\lambda$ are the source location in the ecliptic coordinate system. $\psi$ is the polarization angle.
$\theta_s$ and $\phi_s$ are the ecliptic coordinates of the reference source.
For the reference source of TianQin is J0806, $\theta_s=-4.7040^\circ$ and $\phi_s=120.4425^\circ$.

By performing the same process as described in Section \ref{sec:snr}, we get some expressions similar to Eqs.~(\ref{eq:ave_amp})-(\ref{eq:ave_f}), given below:

\begin{eqnarray}
\langle A^2\rangle &=& \mathcal{A}^2 \left[ (1+\cos^2\iota)^2 \label{eq:ave_amp2}\langle F_{+}^2\rangle+4\cos^2\iota\langle F_{\times}^2\rangle \right] \, ,\\
\langle F_{+}^2\rangle &=& \frac{1}{4}(\cos^22\psi \langle D^2_+ \rangle-\sin4\psi \langle D_+D_\times \rangle + \sin^22\psi \langle D^2_\times \rangle)  \, ,\\
\langle F_{\times}^2\rangle &=& \frac{1}{4}(\cos^22\psi \langle D^2_\times \rangle+\sin4\psi \langle D_+D_\times \rangle + \sin^22\psi \langle D^2_+ \rangle) \, .
\end{eqnarray}
where
\begin{eqnarray}
\langle D^2_+ \rangle &=& b^2_1+b^2_2  \, ,\nn\\
\langle D^2_\times \rangle &=& b^2_3+b^2_4  \, ,\nn\\
\langle D_+D_\times \rangle &=& -2(b_1b_3+b_2b_4)  \, ,
\end{eqnarray}
and
\begin{eqnarray}
b_1 &=& \frac{\sqrt{3}}{8}((3+\cos2\theta)\sin\theta_s\sin2(\phi-\phi_s)+2\sin(\phi-\phi_s)\sin2\theta\cos\theta_s)\, ,\nn\\
b_2 &=& \frac{\sqrt{3}}{32}(3+\cos2(\phi-\phi_s)(9+\cos2\theta(3-\cos2\theta_s))+6\cos2\theta_s\sin^2(\phi-\phi_s) \nn\\
& &-6\cos2\theta\cos^2\theta_s+4\cos(\phi-\phi_s)\sin2\theta\sin2\theta_s) \, ,\nn\\
b_3 &=& \frac{\sqrt{3}}{2}(\cos2(\phi-\phi_s)\cos\theta\sin\theta_s+\cos(\phi-\phi_s)\sin\theta\cos\theta_s)\, ,\nn\\
b_4 &=& \frac{\sqrt{3}}{8}((3-\cos2\theta_s)\cos\theta\sin2(\phi_s-\phi)+2\sin(\phi_s-\phi)\sin\theta\sin2\theta_s) \label{eq:b}\, .
\end{eqnarray}

The average amplitude calculated by Eqs.~(\ref{eq:ave_amp2})-(\ref{eq:b}) is consistent with Eqs.~(\ref{eq:ave_amp})-(\ref{eq:ave_f}), with $0.1\%\sim1\%$ of relative uncertainty.

\section{Coordinate transformation}\label{Coordinate transformation}
The transformation of the source position from the ecliptic coordinates ($\beta,\lambda$) to the detector coordinates ($\theta_S,\phi_S$) and ($\theta_S',\phi_S'$) of the TianQin (TQ) and TQ II is described by the following formula:
\begin{equation}      
\left(                 
\begin{array}{c}   
d\sin\theta_S\cos\phi_S \\  
d\sin\theta_S\sin\phi_S \\ 
d\cos\theta_S \\  
\end{array}
\right) =
R_x(\theta = 120^\circ-90^\circ)R_z(\theta = -4.7^\circ-90^\circ)  
\left(                 
\begin{array}{c}   
d\cos\beta\cos\lambda \\  
d\cos\beta\sin\lambda \\ 
d\sin\beta            \\  
\end{array}
\right)       
\end{equation}
\begin{equation}
{\rm and}~
\left(                 
\begin{array}{c}   
d\sin\theta_S'\cos\phi_S' \\  
d\sin\theta_S'\sin\phi_S' \\ 
d\cos\theta_S' \\ 
\end{array}
\right) =
R_y(\theta = 90^\circ)R_x(\theta = 120^\circ-90^\circ)R_z(\theta = -4.7^\circ-90^\circ)  
\left(                 
\begin{array}{c}   
d\cos\beta\cos\lambda \\  
d\cos\beta\sin\lambda \\ 
d\sin\beta            \\   
\end{array}
\right),
\end{equation}
where the rotation matrices are
\begin{equation} 
R_x(\theta) =    
\left(                
\begin{array}{ccc}  
1 & 0 & 0\\ 
0 & \cos\theta & \sin\theta\\ 
0 & -\sin\theta & \cos\theta\\ 
\end{array}
\right),
R_y(\theta) =    
\left(                
\begin{array}{ccc}  
\cos\theta & 0 & -\sin\theta\\ 
0 & 1 & 0\\ 
\sin\theta & 0 & \cos\theta\\ 
\end{array}
\right) {\rm, and}~ 
R_z(\theta) =    
\left(                
\begin{array}{ccc}  
\cos\theta & \sin\theta & 0\\ 
-\sin\theta & \cos\theta & 0\\ 
0 & 0 & 1\\ 
\end{array}
\right).                 
\end{equation}
\end{widetext}

\bibliographystyle{apsrev4-1}
\bibliography{GDWDs}
\end{document}